\documentclass[10pt]{article}

\usepackage{citesort}
\usepackage{a4wide}
\usepackage{amsmath,amsthm,amsfonts,amssymb}

\makeatletter
\@addtoreset{equation}{section}

\makeatother

\def\e{\mathrm{e}}

\def\Define{:=}
\def\definE{=:}
\def\nn{\nonumber}
\def\id{\mathrm{id}}
\def\End{\mathrm{End}}

\def\tr{\mathrm{tr}}
\renewcommand{\leq}{\leqslant}
\renewcommand{\geq}{\geqslant}
\newtheorem{thm}{Theorem}[section]
\newtheorem{prop}[thm]{Proposition}
\newtheorem{lm}[thm]{Lemma}
\newtheorem{cor}[thm]{Corollary}
\newtheorem{defn}[thm]{Definition}
\newtheorem{conj}[thm]{Conjecture}

\title{\bf An algebraic derivation 
of the eigenspaces associated with an Ising-like spectrum 
of the superintegrable chiral Potts model 
}

\author{
Akinori Nishino\footnote{E-mail address: %
nishino@iis.u-tokyo.ac.jp}
and
Tetsuo Deguchi$^{1}$\footnote{E-mail address: %
deguchi@phys.ocha.ac.jp}\\
\,\\
Institute of Industrial Science, The University of Tokyo, \\
4--6--1 Komaba, Meguro-ku, Tokyo, 153--8505, Japan \\
\,\\
$^{1}$Department of Physics, Ochanomizu University,\\
2--1--1 Otsuka, Bunkyo-ku, Tokyo, 112--8610, Japan
}

\date{\today}

\begin{document}

\setlength{\baselineskip}{16pt}
\def\arraystretch{1.5}

\maketitle

\begin{abstract}
In terms of the $\mathfrak{sl}_{2}$ loop algebra and 
the algebraic Bethe-ansatz method, we derive the invariant subspace 
associated with a given Ising-like spectrum 
consisting of $2^{r}$ eigenvalues 
of the diagonal-to-diagonal transfer 
matrix of the superintegrable chiral Potts (SCP) model  
with arbitrary inhomogeneous parameters. 
We show that every regular Bethe eigenstate of the $\tau_2$-model 
leads to an Ising-like spectrum and is  
an eigenvector of the SCP transfer matrix which is given by 
the product of two diagonal-to-diagonal transfer matrices with 
a constraint on the spectral parameters. 
We also show in a sector that the $\tau_2$-model commutes 
with the $\mathfrak{sl}_{2}$ loop algebra, 
$L(\mathfrak{sl}_{2})$,  
and every regular Bethe state of the $\tau_2$-model is of highest weight. 
Thus, from physical assumptions such as 
the completeness of the Bethe ansatz, 
it follows in the sector 
that every regular Bethe state of the $\tau_2$-model 
generates an $L(\mathfrak{sl}_{2})$-degenerate eigenspace and it  
gives the invariant subspace, i.e. the direct sum of the eigenspaces associated
 with the Ising-like spectrum.  

\end{abstract}
\mathversion{bold}
\section{Introduction}
\mathversion{normal}

The chiral Potts model~\cite{Howes-Kadanoff-Nijs_83NPB,%
vonGehlen-Rittenberg_85NPB,AuYang-McCoy-Perk-Tang-Yan_87PLA,%
Baxter-Perk-AuYang_88PLA,AuYang-Perk_89ASPM},
which is an $N$-state generalization of
the two-dimensional Ising model, has been extensively 
studied from various points of view in recent years.
The model is solvable 
in the sense that its Boltzmann weights
satisfy the star-triangle relation to give a commutative 
family of transfer matrices~\cite{Baxter-Perk-AuYang_88PLA}.
In fact, the free energy, interfacial tension
and order parameters of the model are exactly calculated 
in the thermodynamic limit~\cite{Albertini-McCoy-Perk-Tang_89NPB,%
Baxter_93JSP,Baxter_03PhysicaA,Baxter_06PRL}.

In the superintegrable case of the chiral Potts model,
all the eigenvalues of the transfer matrix are grouped into 
sets of $2^{r}$ eigenvalues similar to those of free fermions.  
We call it the superintegrable chiral Potts (SCP) model 
and the set of eigenvalues an Ising-like spectrum~\cite{%
Albertini-McCoy-Perk-Tang_89NPB,Albertini-McCoy-Perk_89ASPM,%
Baxter_88PLA,Baxter_89JSP,Baxter_93JSP,Tarasov_90PLA}.
The Onsager algebra is powerful to derive 
the spectrum of the two-dimensional Ising 
model~\cite{Onsager_44PR,Perk_89PSPM,Davies_90JPA}, 
in which a set of $2^{r}$ eigenvalues corresponds to 
a $2^{r}$-dimensional irreducible representation of the algebra.
The approach is extended to the $\mathbb{Z}_{N}$-symmetric 
quantum system corresponding to 
the SCP model~\cite{vonGehlen-Rittenberg_85NPB,%
Davies_90JPA,Date-Roan_00CJP}.
However, in contrast to the Ising-case,
the approach does not work enough to derive 
an exact form of the spectrum for $N \geq 3$.  
A derivation of the exact form is established 
by the approach~\cite{Albertini-McCoy-Perk_89ASPM,%
Baxter_88PLA,Baxter_89JSP,Baxter_93JSP,Tarasov_90PLA} 
using functional relations 
among diagonal-to-diagonal transfer 
matrices of the SCP model~\cite{Baxter-Bazhanov-Perk_90IJMPB}.
There, the Ising-like spectrum is described
by a polynomial, which we call the SCP polynomial.
However, it is still nontrivial to define  the SCP polynomial 
by an algebraic method.

In this paper, we present a method for constructing basis vectors 
of the direct sum of the eigenspaces 
associated with a given Ising-like spectrum 
of the transfer matrix of the SCP model in some sector. 
In short, we construct the 
invariant subspace of the Ising-like spectrum. 
First, by the algebraic Bethe-ansatz method,  
we show that every regular Bethe state of the $\tau_{2}$-model 
is an eigenvector of the SCP transfer matrix. Here it is  
defined by the product of two diagonal-to-diagonal 
transfer matrices of the SCP model 
with a constraint on the spectral parameters. 
We shall define it in detail in Section~\ref{sec:def_SCP}. 
The $\tau_{2}$-model is the integrable $N$-state spin chain 
corresponding to a nilpotent case of the cyclic 
$L$-operator~\cite{Korepanov_95SPMJ,Korepanov_97JMS}; 
the transfer matrix constructed from the cyclic $L$-operators 
commutes with the transfer matrix of the chiral Potts 
model~\cite{Bazhanov-Stroganov_90NPB}. 
Secondly, we show in a sector that the $\tau_{2}$-model 
has the symmetry of the $\mathfrak{sl}_{2}$ loop algebra, 
$L(\mathfrak{sl}_{2})$, and also in the sector that 
every regular Bethe state of the $\tau_{2}$-model 
is a highest weight vector of the symmetry. 
Thus, the degenerate eigenspaces are generated by regular 
Bethe eigenstates~\cite{Nishino-Deguchi_06PLA,Deguchi_07JPA} 
in the sector through the symmetry. 
Here we shall define regular Bethe states 
in Section~\ref{sec:def_tau-2-model}.  
Thirdly, with some physical assumptions 
such as the completeness of the Bethe ansatz, 
we show that for the diagonal-to-diagonal transfer matrix 
of the SCP model the invariant subspace of the Ising-like spectrum 
associated with a regular Bethe state is 
given by the $L(\mathfrak{sl}_{2})$-degenerate eigenspace 
of the $\tau_{2}$-model generated by the same regular Bethe state. 

We apply a generalization of the algebraic Bethe ansatz to the 
SCP transfer matrix with arbitrary inhomogeneous parameters, 
and do not use the functional relations 
among the transfer matrices~\cite{Baxter-Bazhanov-Perk_90IJMPB}.
The algebraic approach treats the SCP model and the $\tau_{2}$-model 
in a unified way, which might be useful for calculating  
correlation functions for the model.

We reproduce the SCP polynomial as a kind of Drinfeld polynomial 
which characterizes the finite-dimensional highest weight representation 
of $L(\mathfrak{sl}_{2})$ generated by the regular Bethe state. 
Here it is not necessarily irreducible \cite{Deguchi_07JSM}. 
For generic values of inhomogeneous parameters, however,  
the zeros of the polynomial should be distinct, and hence 
the highest weight representation should be irreducible.
Thus, the SCP polynomial should be identified with the Drinfeld 
polynomial~\cite{Chari-Pressley_91CMP,Drinfeld_88SMD,%
Deguchi_06DGP,Deguchi_07JSM}. 

The algebraic derivation of the invariant subspace associated 
with the Ising-like spectrum 
proves in the sector a previous conjecture that 
for the $\mathbb{Z}_N$-symmetric Hamiltonian 
the representation space of the Onsager algebra associated with 
an SCP polynomial should correspond to the  
$L(\mathfrak{sl}_{2})$-degenerate eigenspace of the $\tau_2$-model 
associated with the Drinfeld polynomial~\cite{Nishino-Deguchi_06PLA}. 

For the $\tau_{2}$-model we shall show a Borel subalgebra symmetry 
of $L(\mathfrak{sl}_{2})$ 
through a gauge transformation on the $L$-operators.  
In fact, it is known that every finite-dimensional 
irreducible representation 
of the Borel subalgebra is extended to that of the 
$\mathfrak{sl}_{2}$ loop algebra~\cite{%
Benkart-Terwilliger_04JA,Deguchi_07math-phys}.
We shall thus derive the $L(\mathfrak{sl}_{2})$ symmetry 
of the $\tau_{2}$-model in the general $N$-state case 
with inhomogeneous parameters in the paper. Previously,   
the symmetry has been shown for the odd $N$ and 
homogeneous case~\cite{Nishino-Deguchi_06PLA}. 
The present result also proves 
the $L(\mathfrak{sl}_{2})$ symmetry of the $\tau_{2}$-model 
for the even $N$ and homogeneous case. It thus proves 
the conjecture \cite{AuYang-Perk_08JPA} 
that the proposed set of commuting operators 
forms the $L(\mathfrak{sl}_{2})$ symmetry 
of  the $\tau_{2}$-model.  
The $L(\mathfrak{sl}_{2})$ symmetry of 
the $\tau_{2}$-model is closely related to  
that of the spin-1/2 XXZ spin chain 
at roots of unity~\cite{Deguchi-Fabricius-McCoy_01JSP,%
Deguchi_04JPA,Deguchi_07JSM,Deguchi_07JPA}. 

The article consists of the following:
in Section 2, we introduce the SCP model and 
the $\tau_{2}$-model.
We review the algebraic Bethe-ansatz method
for the $\tau_{2}$-model~\cite{Tarasov_90PLA} and 
the Yang-Baxter relation between the monodromy matrices
of the two models~\cite{Bazhanov-Stroganov_90NPB}.
In Section 3, generalizing 
the algebraic Bethe ansatz, we show that 
every regular Bethe eigenstate of the $\tau_{2}$-model is 
an eigenvector of the SCP transfer matrix with a constraint 
on the spectral parameters. 
The expression of eigenvalues of the product 
of two diagonal-to-diagonal transfer matrices suggests 
the Ising-like spectrum to each of the two transfer matrices.
In Section 4, 
we show in a sector the Borel subalgebra symmetry 
of the $\tau_{2}$-model through a gauge transformation 
on the $L$-operators. 
It thus follows from 
\cite{Benkart-Terwilliger_04JA,Deguchi_07math-phys}
that the $\tau_{2}$-model has the $L(\mathfrak{sl}_{2})$ symmetry.   
We also show in the sector that every regular Bethe state 
of the $\tau_{2}$-model generates 
an irreducible highest weight representation of 
$L(\mathfrak{sl}_{2})$, 
which gives the degenerate eigenspace associated 
with the regular Bethe state 
for the $\tau_{2}$-model. 
We then formulate the conjecture 
that the diagonal-to-diagonal transfer matrix 
of the SCP model has the Ising-like spectrum in the 
$L(\mathfrak{sl}_{2})$-degenerate 
eigenspace of the $\tau_{2}$-model.

\mathversion{bold}
\section{Models and Yang-Baxter relations}
\mathversion{normal}

\mathversion{bold}
\subsection{The  chiral Potts model
and the superintegrable conditions 
}
\mathversion{normal}
\label{sec:def_SCP}

We briefly review the chiral Potts 
model~\cite{AuYang-McCoy-Perk-Tang-Yan_87PLA,AuYang-Perk_89ASPM,%
Baxter-Perk-AuYang_88PLA}
and its superintegrable point~\cite{Albertini-McCoy-Perk-Tang_89NPB,%
Albertini-McCoy-Perk_89ASPM,Baxter_88PLA,Baxter_89JSP,Baxter_93JSP}.
The model is defined on a two-dimensional square lattice 
with $N$-state local spins interacting along the edges. 
For two adjacent local spins $\sigma_{i}$ and $\sigma_{j}$ 
which take values in $\mathbb{Z}_{N}$, 
that is, $0$, $1$, \ldots, $N-1$, 
two edge-types of the Boltzmann weights 
$W_{pq}(\sigma_{i}-\sigma_{j})$
and $\Bar{W}_{pq}(\sigma_{i}-\sigma_{j})$ are given as
\begin{align}
&W_{pq}(n)
 =W_{pq}(0)\prod_{j=1}^{n}\frac{\mu_{p}}{\mu_{q}}
  \frac{y_{q}-x_{p}\omega^{j}}{y_{p}-x_{q}\omega^{j}},
 \qquad
 \Bar{W}_{pq}(n)
 =\Bar{W}_{pq}(0)\prod_{j=1}^{n}\mu_{p}\mu_{q}
  \frac{x_{p}\omega-x_{q}\omega^{j}}{y_{q}-y_{p}\omega^{j}},
 \nn
\end{align}
where $\omega$ is an $N$th root of unity. Here 
$p=(x_{p}, y_{p}, \mu_{p})$ and $q=(x_{q}, y_{q}, \mu_{q})$, 
which we call rapidities, are given on a Fermat curve defined by
\begin{align}
\label{eq:Fermat}
 kx_{p}^{N}=1-k^{\prime}\mu_{p}^{-N},\quad
 ky_{p}^{N}=1-k^{\prime}\mu_{p}^{N},\quad
 k^{2}+k^{\prime 2}=1.
\end{align}
Note that both $W_{pq}(n)$ and $\Bar{W}_{pq}(n)$
are functions of variable $n\in\mathbb{Z}_{N}$.
The model is integrable in the sense 
that the Boltzmann weights satisfy the star-triangle 
relations~\cite{Baxter-Perk-AuYang_88PLA,AuYang-Perk_89ASPM}.
We also give the Fourier-transformed Boltzmann weight:
\begin{align}
&\widehat{W}_{pq}(n)
 =\sum_{m=0}^{N-1}\omega^{-nm}W_{pq}(m)
 =\widehat{W}_{pq}(0)\prod_{j=1}^{n}
  \frac{x_{p}\mu_{p}\omega-x_{q}\mu_{q}\omega^{j}}
       {y_{q}\mu_{p}-y_{p}\mu_{q}\omega^{j}}.
 \nn
\end{align}

We introduce the $S$-operator~\cite{Bazhanov-Stroganov_90NPB,%
Date-Jimbo-Miki-Miwa_CMP91} to construct the monodromy matrix
of the SCP model.
Let $Z$ and $X$ be operators which have the action
$Zv_{\sigma}=\omega^{\sigma}v_{\sigma}$
and $Xv_{\sigma}=v_{\sigma+1}$
for a standard basis $\{v_{\sigma}|\sigma\in\mathbb{Z}_{N}\}$
of the $N$-dimensional vector space $\mathbb{C}^{N}$.
By using them and combining the Boltzmann weights,
we define the $S$-operator 
$S(p,p^{\prime};q,q^{\prime})\in\End\big(
\mathbb{C}^{N}\otimes\mathbb{C}^{N}\big)$ by
\begin{align}
\label{eq:S-op_CP}
&S(p,p^{\prime};q,q^{\prime})
 =\frac{1}{N^{2}}P_{\mathbb{C}^{N}\otimes\mathbb{C}^{N}}
  \sum_{\{n_{i}\}}
  w_{pp^{\prime}qq^{\prime}}(n_{1},n_{2},n_{3},n_{4})
  X^{n_{1}}Z^{n_{2}}X^{n_{4}}\otimes
  X^{-n_{1}}Z^{n_{3}}X^{-n_{4}},
 \\
&w_{pp^{\prime}qq^{\prime}}(n_{1},n_{2},n_{3},n_{4})
=\frac{\widehat{W}_{pq^{\prime}}(n_{1})}
      {W_{pq^{\prime}}(0)}
 \frac{\Bar{W}_{pq}(n_{2})}
      {\Bar{W}_{pq}(0)}
 \frac{\Bar{W}_{p^{\prime}q^{\prime}}(n_{3})}
      {\Bar{W}_{p^{\prime}q^{\prime}}(0)}
 \frac{\widehat{W}_{p^{\prime}q}(n_{4})}
      {W_{p^{\prime}q}(0)},
 \nn
\end{align}
where $P_{\mathbb{C}^{N}\otimes\mathbb{C}^{N}}$ 
is the standard permutation operator:
$P_{\mathbb{C}^{N}\otimes\mathbb{C}^{N}}:
v_{\sigma}\otimes v_{\tau}
\mapsto v_{\tau}\otimes v_{\sigma}$.
The action of the $S$-operator is extended to a tensor product
$(\mathbb{C}^{N})^{\otimes L}\otimes\mathbb{C}^{N}$,
where the tensor product $(\mathbb{C}^{N})^{\otimes L}$
is the quantum space describing an $L$-site spin chain
and the last space $\mathbb{C}^{N}$ is an auxiliary space.
We denote by $S_{i}(p,p^{\prime};q,q^{\prime})$
the $S$-operator acting on the $i$th component of
$(\mathbb{C}^{N})^{\otimes L}$ and auxiliary space 
$\mathbb{C}^{N}$ as the $S$-operator 
$S(p,p^{\prime};q,q^{\prime})$
and other components of $(\mathbb{C}^{N})^{\otimes L}$
as the identity. 
Here we use the operators $Z_{i}$ and $X_{i}$ 
on $(\mathbb{C}^{N})^{\otimes L}$ given by
\[
 Z_{i}=\id\otimes\cdots\otimes \overset{i}{\Check{Z}}
 \otimes\cdots\otimes\id,\qquad
 X_{i}=\id\otimes\cdots\otimes \overset{i}{\Check{X}}
 \otimes\cdots\otimes\id.
\]
We construct  monodromy matrix
$T(q_{1},q_{2};\{p,p^{\prime}\})\in\End\big(
(\mathbb{C}^{N})^{\otimes L}\otimes\mathbb{C}^{N}\big)$
and transfer matrix 
$t(q_{1},q_{2};\{p,p^{\prime}\})\in\End\big(
(\mathbb{C}^{N})^{\otimes L}\big)$ as
\begin{align}
\label{eq:transfer-matrix_SCP}
&T(q_{1},q_{2};\{p,p^{\prime}\})=
 \prod_{i=1}^{L}
 S_{i}(p_{i},p^{\prime}_{i};q_{1},q_{2}),\qquad
 t(q_{1},q_{2};\{p,p^{\prime}\})=
 \tr_{\mathbb{C}^{N}}\big(
 T(q_{1},q_{2};\{p,p^{\prime}\})\big),
\end{align}
where both $p_{i}$ and $p^{\prime}_{i}$ are rapidities 
of the $i$th component of the quantum space 
$(\mathbb{C}^{N})^{\otimes L}$ and $q_{1}$ and $q_{2}$ are those
of the auxiliary space $\mathbb{C}^{N}$.
The parameters $q_{1}$ and $q_{2}$ are called spectral parameters.
Here the symbol $\{p, p^{\prime} \}$ denotes 
the set of rapidities $p_{i}$ and $p_{i}^{\prime}$ for 
$i=1, 2, \ldots, L$. 

The transfer matrices satisfy the commutativity 
\[
t(q_{1},q_{2};\{p,p^{\prime}\})t(r_{1},r_{2};\{p,p^{\prime}\})
=t(r_{1},r_{2};\{p,p^{\prime}\})t(q_{1},q_{2};\{p,p^{\prime}\}),
\]
which is a result of the star-triangle 
relation~\cite{Baxter-Perk-AuYang_88PLA}.
Then the eigenvectors of the transfer matrix
$t(q_{1},q_{2};\{p,p^{\prime}\})$
are independent of the spectral parameters $q_{1}$ or $q_{2}$.

We call the transfer matrix $t(q_{1},q_{2};\{p,p^{\prime}\})$  
the row-to-row transfer matrix of the chiral Potts model
since the Boltzmann weight 
$w_{pp^{\prime}qq^{\prime}}(n_{1},n_{2},n_{3},n_{4})$
is considered as that of a vertex model.
The row-to-row transfer matrix $t(q_{1},q_{2};\{p,p^{\prime}\})$
is given by  the product of two types of diagonal-to-diagonal 
transfer matrices $T_{\text{D}}(x_{q_{1}},y_{q_{1}})$
and $\Hat{T}_{\text{D}}(x_{q_{2}},y_{q_{2}})$ 
which are defined by 
\begin{align}
& T_{\text{D}}(x_{q},y_{q})_{\sigma}^{\sigma^{\prime}}
 =\prod_{i=1}^{L}
  \frac{W_{p^{\prime}_{i}q}(\sigma_{i}-\sigma_{i}^{\prime})}
  {W_{p^{\prime}_{i}q}(0)}
  \frac{\Bar{W}_{p_{i+1}q}(\sigma_{i+1}-\sigma_{i}^{\prime})}
  {\Bar{W}_{p_{i+1}q}(0)},
  \nn\\
& \Hat{T}_{\text{D}}(x_{q},y_{q})_{\sigma}^{\sigma^{\prime}}
 =\prod_{i=1}^{L}
  \frac{\Bar{W}_{p^{\prime}_{i}q}(\sigma_{i}-\sigma_{i}^{\prime})}
  {\Bar{W}_{p^{\prime}_{i}q}(0)}
  \frac{W_{p_{i+1}q}(\sigma_{i}-\sigma_{i+1}^{\prime})}
  {W_{p_{i+1}q}(0)},
  \nn
\end{align}
where the periodic boundary conditions $\sigma_{L+1}=\sigma_{1}$
and $p_{L+1}=p_{1}$ are imposed.
The diagonal-to-diagonal transfer matrices are
diagonalized by a pair of invertible matrices
$U$ and $V$, which are independent of the parameter $q$, as
$U^{-1}T_{\text{D}}(x_{q},y_{q})V=\Lambda(x_{q},y_{q})$
and $V^{-1}\Hat{T}_{\text{D}}(x_{q},y_{q})U
=\Hat{\Lambda}(x_{q},y_{q})$.

Let us now discuss the superintegrable case. 
When rapidities $p$ and $p^{\prime}$ satisfy the conditions 
$x_{p}=y_{p^{\prime}}, y_{p}=x_{p^{\prime}}, 
\mu_{p}=\mu_{p^{\prime}}^{-1}$, 
we denote the rapidity $p^{\prime}$ by $\bar p$.   

\begin{defn}
We call the chiral Potts model superintegrable,  
if rapidities $\{p, p^{\prime} \}$ 
satisfy the conditions $p^{\prime}_{i}= {\bar p}_i$ for all $i$, 
that is, $x_{p_{i}}=y_{p^{\prime}_{i}}, y_{p_{i}}=x_{p^{\prime}_{i}}, 
\mu_{p_{i}}=\mu_{p^{\prime}_{i}}^{-1},(1\leq i\leq L)$
~\cite{Baxter_88PLA,Baxter_89JSP,Baxter_93JSP}.
\end{defn}

In the superintegrable case, 
we denote by $T(q_{1},q_{2};\{p\})$ and $t(q_{1},q_{2};\{p\})$
 the monodromy matrix and the row-to-row transfer matrix of the 
chiral Potts model, respectively.  
That is, we express 
$T(q_{1},q_{2};\{p\})=T(q_{1},q_{2};\{p, {\bar p} \})$ and 
$t(q_{1},q_{2};\{p\})=t(q_{1},q_{2};\{p, {\bar p} \})$. 
Hereafter we call the row-to-row transfer matrix $t(q_{1},q_{2};\{p\})$ 
the SCP transfer matrix, briefly. We also call 
$T(q_{1},q_{2};\{p\})$ the SCP monodromy matrix. 

\mathversion{bold}
\subsection{The $\tau_{2}$-model and the algebraic Bethe ansatz}
\mathversion{normal}
\label{sec:def_tau-2-model}

Let us now introduce an integrable $N$-state spin chain 
whose transfer matrix commutes with the SCP transfer matrix.
We introduce the cyclic $L$-operator
$\mathcal{L}(z;p,p^{\prime})
\in\End(\mathbb{C}^{2}\otimes\mathbb{C}^{N})$~\cite{%
Korepanov_95SPMJ,Korepanov_97JMS} by  
\begin{align}
\label{eq:L-op_CP}
 \mathcal{L}(z;p,p^{\prime})=
 \begin{pmatrix}
  -y_{p}y_{p^{\prime}}z+\mu_{p}\mu_{p^{\prime}}Z
 &-z(y_{p}-x_{p^{\prime}}\mu_{p}\mu_{p^{\prime}}Z)X \\
  X^{-1}(y_{p^{\prime}}-x_{p}\mu_{p}\mu_{p^{\prime}}Z)
 &1-x_{p}x_{p^{\prime}}\mu_{p}\mu_{p^{\prime}}z\omega Z
 \end{pmatrix}. 
\end{align}
In the same way as the $S$-operator,
the action of the $L$-operator 
$\mathcal{L}(z;p,p^{\prime})$ is extended to the tensor product
$\mathbb{C}^{2}\otimes(\mathbb{C}^{N})^{\otimes L}$
where the space $\mathbb{C}^{2}$ is another auxiliary space;
we denote by $\mathcal{L}_{i}(z;p,p^{\prime})$
the $L$-operator acting on the auxiliary space
and $i$th component of 
the quantum space $(\mathbb{C}^{N})^{\otimes L}$
as the $L$-operator $\mathcal{L}(z;p,p^{\prime})$
and other components of $(\mathbb{C}^{N})^{\otimes L}$
as the identity.
The following properties give the reason why the chiral Potts model
is considered as a descendant of the six-vertex 
model~\cite{Bazhanov-Stroganov_90NPB}:

\begin{prop}
The $L$-operators 
$\mathcal{L}_{i}(z)=\mathcal{L}_{i}(z;p,p^{\prime})$ satisfy 
a Yang-Baxter relation
\begin{align}
\label{eq:RLL=LLR}
&R(z/w)\big(\mathcal{L}_{i}(z)
            \otimes\id_{\mathbb{C}^{2}}\big)
       \big(\id_{\mathbb{C}^{2}}\otimes
            \mathcal{L}_{i}(w)\big)
 =\big(\id_{\mathbb{C}^{2}}\otimes
       \mathcal{L}_{i}(w)\big)
  \big(\mathcal{L}_{i}(z)
       \otimes\id_{\mathbb{C}^{2}}\big)R(z/w)
\end{align}
with the $R$-matrix defined by
\begin{align}
\label{eq:R-matrix_hom}
 R(z)=
 \begin{pmatrix}
  1-z\omega & 0 & 0 & 0 \\
  0 & \omega(1-z) & (1-\omega)z & 0 \\
  0 & 1-\omega    & 1-z & 0 \\
  0 & 0 & 0 & 1-z\omega \\
 \end{pmatrix}
 ,
\end{align}
and another Yang-Baxter relation
\begin{align}
\label{eq:SLL-relation_CP}
&S_{ij}(p,p^{\prime};q,q^{\prime})
 \mathcal{L}_{i}(z;p,p^{\prime})
 \mathcal{L}_{j}(z;q,q^{\prime})
=\mathcal{L}_{j}(z;q,q^{\prime})
 \mathcal{L}_{i}(z;p,p^{\prime})
 S_{ij}(p,p^{\prime};q,q^{\prime}),
\end{align}
where $S_{ij}(p,p^{\prime};q,q^{\prime})$
is the $S$-operator acting on the $i$th and $j$th components of 
the quantum space $(\mathbb{C}^{N})^{\otimes L}$
as the $S$-operator $S(p,p^{\prime};q,q^{\prime})$
and other components of $(\mathbb{C}^{N})^{\otimes L}$
as the identity.
\end{prop}

At the superintegrable point, the cyclic $L$-operator 
$\mathcal{L}_{i}(z;p_{i},p^{\prime}_{i})$ is reduced to 
\begin{align}
\label{eq:L-op_SCP}
 \mathcal{L}_{i}(z;p_{i},\Bar{p}_{i})=
 \begin{pmatrix}
  -t_{p_{i}}z+Z_{i} & -y_{p_{i}}z(1-Z_{i})X_{i} \\
  x_{p_{i}}X^{-1}_{i}(1-Z_{i}) & 1-t_{p_{i}}z\omega Z_{i}
 \end{pmatrix},
\end{align}
where we have defined $t_{p}=x_{p}y_{p}$.
We introduce the monodromy matrix 
$\mathcal{T}(z;\{p\})\in\End\big(\mathbb{C}^{2}
\otimes(\mathbb{C}^{N})^{\otimes L}\big)$
and the transfer matrix $\tau(z;\{p\})\in\End\big(
(\mathbb{C}^{N})^{\otimes L}\big)$ by
\begin{align}
\label{eq:T-op_tau-2}
&\mathcal{T}(z;\{p\})=
 \prod_{i=1}^{L}\mathcal{L}_{i}(z;p_{i},\Bar{p}_{i})
 \definE
 \begin{pmatrix}
  A(z) & B(z) \\
  C(z) & D(z) 
 \end{pmatrix}
 ,\quad
 \tau(z;\{p\})=
 \tr_{\mathbb{C}^{2}}\big(\mathcal{T}(z;\{p\})\big).
\end{align}
Here we have also defined operators 
$A(z), B(z), C(z), D(z)\in
\End\big((\mathbb{C}^{N})^{\otimes L}\big)$.
The spin chain described by the transfer matrix $\tau(z;\{p\})$ 
is called the $\tau_{2}$-model~\cite{%
Baxter_06PRL,AuYang-Perk_08JPA,Roan_07JSM}.
We remark that the original $\tau_{2}$-model is defined
in terms of the cyclic $L$-operator 
$\mathcal{L}(z;p,p^{\prime})$~\eqref{eq:L-op_CP}. 

\begin{prop}
The monodromy matrix $\mathcal{T}(z;\{p\})$ 
satisfies a Yang-Baxter relation
\begin{align}
\label{eq:RTT=TTR}
&R(z/w)\big(\mathcal{T}(z;\{p\})\otimes\id_{\mathbb{C}^{2}}\big)
       \big(\id_{\mathbb{C}^{2}}\otimes\mathcal{T}(w;\{p\})\big)
 \nn\\
&=\big(\id_{\mathbb{C}^{2}}\otimes\mathcal{T}(w;\{p\})\big)
  \big(\mathcal{T}(z;\{p\})\otimes\id_{\mathbb{C}^{2}}\big)R(z/w),
\end{align}
where $R(z)$ is the $R$-matrix defined 
in \eqref{eq:R-matrix_hom}.
\end{prop}

The Yang-Baxter relation~\eqref{eq:RTT=TTR} gives
the commutativity 
$\tau(z;\{p\})\tau(w;\{p\})=\tau(w;\{p\})\tau(z;\{p\})$.
Hence the eigenvectors of the transfer matrix $\tau(z;\{p\})$
are independent of the parameter $z$.
The relation also produces relations among operators 
$A(z), B(z), C(z)$ and $D(z)$~\cite{Faddeev-Takhtadzhyan_84JSM}.
In the next section, we need more general relations, 
which are collected in Lemma~\ref{lm:ABCD}.
By using the relations,
the algebraic Bethe-ansatz method is readily applicable to 
the transfer matrix $\tau(z;\{p\})$~\cite{Tarasov_90PLA}.

Let $|0\rangle$ be the reference state 
$v_{0}\otimes v_{0}\otimes\cdots\otimes v_{0}$. 
It has the following properties:
\[
 A(z)|0\rangle=a(z)|0\rangle
 =\prod_{n=1}^{L}(1-t_{p_{n}}z)|0\rangle,
 \quad
 D(z)|0\rangle=d(z)|0\rangle
 =\prod_{n=1}^{L}(1-t_{p_{n}}z\omega)|0\rangle,
 \quad
 C(z)|0\rangle=0,
\]
for arbitrary $z$.

\begin{prop}
\label{prop:ABA-BSK}
Let $\{z_{i}|i=1,2,\ldots, M\}$ be a solution 
of the Bethe equations: 
\begin{align}
\label{eq:Bethe-eq}
 a(z_{i})\prod_{j=1 \atop j(\neq i)}^{M}f(z_{i}/z_{j})
 =d(z_{i})\prod_{j=1 \atop j(\neq i)}^{M}f(z_{j}/z_{i}),
\end{align}
where $f(z)=(z-\omega)/(z-1)\omega$.
Then, vector
$|M \rangle=B(z_{1})B(z_{2})\cdots B(z_{M})|0\rangle$
gives an eigenvector of the transfer matrix 
$\tau(z;\{p\})$:
\begin{align}
\label{eq:tau-eigenvalue_tau-2}
&\tau(z;\{p\})|M \rangle
 =\Big(a(z)\prod_{i=1}^{M}\omega f(z/z_{i})
 +d(z)\prod_{i=1}^{M}\omega f(z_{i}/z)\Big)
 |M \rangle.
\end{align}
The vector $|M \rangle$ is referred to as a Bethe state.
\end{prop}

If solutions of the Bethe equations~\eqref{eq:Bethe-eq}
are non-zero, finite and distinct, 
we call them regular \cite{Deguchi_07JPA}. 
If $\{z_{i}|i = 1, 2, \ldots, R\}$ is a regular solution  
of the Bethe equations, 
we call the Bethe state 
$B(z_{1})B(z_{2})\cdots B(z_{R})|0\rangle$
regular, and denote it by $|R\rangle$.  

\subsection{Commutativity of transfer matrices}

As a consequence of the relation~\eqref{eq:SLL-relation_CP},
we obtain the Yang-Baxter relation
between the monodromy matrices $\mathcal{T}(z;\{p\})$
and $T(q_{1},q_{2};\{p\})$, 
by which we shall generalize the algebraic Bethe-ansatz method.

\begin{prop}
The monodromy matrices $\mathcal{T}(z;\{p\})$
and $T(q_{1},q_{2};\{p\})$ satisfy 
\begin{align}
\label{eq:LTT=TTL}
&\mathcal{L}(z;q_{1},q_{2})
 \big(\mathcal{T}(z;\{p\})\otimes\id_{\mathbb{C}^{N}}\big)
 \big(\id_{\mathbb{C}^{2}}\otimes T(q_{1},q_{2};\{p\})\big)
 \nn\\
&=\big(\id_{\mathbb{C}^{2}}\otimes T(q_{1},q_{2};\{p\})\big)
  \big(\mathcal{T}(z;\{p\})\otimes\id_{\mathbb{C}^{N}}\big)
  \mathcal{L}(z;q_{1},q_{2}).
\end{align}
Here the cyclic $L$-operator $\mathcal{L}(z;q_{1},q_{2})$
defined in \eqref{eq:L-op_CP} is considered as 
a $2N\times 2N$ matrix acting on the tensor product
$\mathbb{C}^{2}\otimes\mathbb{C}^{N}$ of the auxiliary spaces.
As a corollary, the transfer matrix $\tau(z;\{p\})$ commutes
with the transfer matrix $t(q_{1},q_{2};\{p\})$.
\end{prop}

Thanks to the commutativity of the two transfer matrices 
$\tau(z;\{p\})$ and $t(q_{1},q_{2};\{p\})$, 
they may have a set of common eigenvectors. 
For the $\tau_{2}$-model, 
we have obtained the eigenstates 
through the algebraic Bethe-ansatz method. 
If a given Bethe eigenvector of the $\tau_{2}$-model 
has a non-degenerate eigenvalue of $\tau(z; \{p\})$,  
then it also becomes an eigenvector of 
the SCP transfer matrix $t(q_{1},q_{2};\{p\})$. 
However, in section 4,  we shall show in a sector that  
the transfer matrix $\tau(z;\{p\})$ of the $\tau_{2}$-model 
has degenerate eigenvectors with respect 
to the $\mathfrak{sl}_{2}$ loop algebra  
and hence not all the Bethe states of the $\tau_{2}$-model 
are necessarily eigenvectors of  $t(q_{1},q_{2};\{p\})$. 

\mathversion{bold}
\section{Spectrum of the superintegrable chiral Potts model}
\mathversion{normal}
\label{sec:Spectrum_SCP}

We shall show in this section that, if the spectral parameters
$q_{1}$ and $q_{2}$ satisfy the condition $q_{2}={\bar q_{1}}(s)
=(y_{q_{1}},x_{q_{1}}\omega^{s},\mu_{q_{1}}^{-1})$,
every regular Bethe eigenstate of $\tau(z;\{p\})$ 
is an eigenstate of the SCP transfer matrix 
$t(q_{1},q_{2};\{p\})$. 

\subsection{Algebraic Bethe-ansatz method for the SCP transfer matrix}

First, we give a fundamental relation,  
generalizing the standard algebraic Bethe-ansatz method.

\begin{prop}
\label{prop:TBBBOmega-relation}
Let $B_{i}$, $A_{i}$ and $D_{i}$ denote 
$B(z_{i})$, $A(z_{i})$ and $D(z_{i})$, respectively.
Let $T_{\tau}^{\tau^{\prime}}$,
$(\tau, \tau^{\prime}\in\mathbb{Z}_{N})$
denote the operator-valued entries of the SCP 
monodromy matrix $T(q_{1},q_{2};\{p\})$.
By setting $q_{1}=q=(x_{q},y_{q},\mu_{q})$ and
$q_{2}=\Bar{q}(s)=(y_{q},x_{q}\omega^{s},\mu_{q}^{-1})$, 
$(s=0,1,\ldots,N-1)$
we have
\begin{align}
\label{eq:TBBBOmega}
&B_{1}\cdots B_{n}
 T^{\tau^{\prime}}_{\tau}
 |0\rangle
 \nn\\
&=\hspace{-10pt}
  \sum_{{\{i_{\ell}\}, \{j_{\ell}\}, \{k_{\ell}\} 
  \atop n_{B}+n_{A}+n_{D}=n}}
  \hspace{-10pt}c^{\tau^{\prime}\tau}_{n}
  (\{i_{\ell}\};\{j_{\ell}\};\{k_{\ell}\})
  a(z_{j_{1}})\cdots a(z_{j_{n_{A}}})
  d(z_{k_{1}})\cdots d(z_{k_{n_{D}}})
  T^{\tau^{\prime}-n_{D}}_{\tau+n_{A}}
  B_{i_{1}}\cdots B_{i_{n_{B}}}
  |0\rangle.
\end{align}
Here $\{i_{\ell}\}, \{j_{\ell}\}$ and $\{k_{\ell}\}$
are such disjoint subsets of 
the index set $\Sigma_{n}=\{1,2,\ldots,n\}$ that the numbers 
of elements of the subsets denoted by $\sharp\{i_{\ell}\}=n_{B}$,
$\sharp\{j_{\ell}\}=n_{A}$ and $\sharp\{k_{\ell}\}=n_{D}$, 
respectively, satisfy the condition $n_{B}+n_{A}+n_{D}=n$, 
and the coefficients 
$c^{\tau^{\prime}\tau}_{n}
 (\{i_{\ell}\};\{j_{\ell}\};\{k_{\ell}\})$ are given by
\begin{align}
&c^{\tau^{\prime}\tau}_{n}
 (\{i_{\ell}\};\{j_{\ell}\};\{k_{\ell}\})
 \nn\\
&=\prod_{\ell=1}^{n_{B}}
  \frac{1}{\mu_{\tau^{\prime}\tau}(z_{i_{\ell}})}
  \prod_{\ell=1}^{n_{A}}
  \frac{\nu_{\tau+\ell}(z_{j_{\ell}})}
  {\mu_{\tau^{\prime},\tau+\ell-1}(z_{j_{\ell}})}
  \prod_{\ell=1}^{n_{D}}
  \frac{-\nu_{\tau^{\prime}-\ell+1}(z_{k_{\ell}})}
  {\mu_{\tau^{\prime}-\ell+1,\tau}(z_{k_{\ell}})}
 \!\!\prod_{i\in\{i_{\ell}\}\atop j\in\{j_{\ell}\}}\!\!
 \omega f_{ji}
 \!\!\prod_{i\in\{i_{\ell}\}\atop k\in\{k_{\ell}\}}\!\!
 \omega f_{ik}
 \!\!\prod_{j\in\{j_{\ell}\}\atop k\in\{k_{\ell}\}}\!\!
 \omega f_{jk},
 \nn
\end{align}
with $\mu_{\tau^{\prime}\tau}(z)$,
$\nu_{\tau}(z)$ and $f_{ij}$ defined by
\begin{align}
\label{eq:mu-nu}
&\mu_{\tau^{\prime}\tau}(z)
 =\frac{(t_{q}z-1)(t_{q}z\omega^{s}-1)\omega^{\tau^{\prime}}}
  {(t_{q}z\omega^{\tau^{\prime}}-1)(t_{q}z\omega^{\tau+1}-1)},
 \quad
 \nu_{\tau}(z)
 =\frac{y_{q}z(1-\omega^{\tau})}{t_{q}z\omega^{\tau}-1},
 \quad
 f_{ij}=\frac{z_{i}-z_{j}\omega}{(z_{i}-z_{j})\omega}.
\end{align}
\end{prop}

A proof of the relation is presented in the next subsection.
The point of the proof is to arrange
the product $B_{1}\cdots B_{n} T^{\tau^{\prime}}_{\tau}$
into the order $TBADC$, which is possible
by the help of the relations \eqref{eq:ABCD-relation} 
and \eqref{eq:TB-relation}.
On the reference state $|0\rangle$,
the terms with the operator $C(z_{i})$ vanish 
and the operators $A(z_{i})$ and $D(z_{i})$
are replaced by
the factors $a(z_{i})$ and $d(z_{i})$, respectively.

Next, we apply the transfer matrix $t(q,\Bar{q}(s);\{p\})$ 
to the reference state $|0\rangle$.
{}It is directly shown that
the reference state $|0\rangle$ is an eigenvector of the transfer 
matrix $t(q,\Bar{q}(s);\{p\})$~\cite{Baxter_88PLA},
\begin{align}
\label{eq:t-eigenvalue-0_SCP}
&t(q,\Bar{q}(s);\{p\})|0\rangle
 =\sum_{\tau=0}^{N-1}\lambda_{\tau}|0\rangle
 \nn\\
&=N^{L}\left(\prod_{n=1}^{L}
 \frac{x_{p_{n}}-x_{q}}{x_{p_{n}}^{N}-x_{q}^{N}}
 \frac{y_{p_{n}}-y_{q}}{y_{p_{n}}^{N}-y_{q}^{N}}\right)
 \sum_{\tau=0}^{N-1}\left(\prod_{n=1}^{L}
 \frac{t_{p_{n}}^{N}-t_{q}^{N}}{t_{p_{n}}-t_{q}\omega^{\tau}}
 \right)\omega^{\tau L}
 |0\rangle,
\end{align}
where $t_{p_{n}}=x_{p_{n}}y_{p_{n}}$ and $t_{q}=x_{q}y_{q}$. 
Here we define $\lambda_{\tau}$ by 
relation \eqref{eq:t-eigenvalue-0_SCP}. 

Let $\{z_{i}|i=1,2,\ldots,R\}$ be a regular solution of the Bethe 
equations~\eqref{eq:Bethe-eq} and extract the term 
$T^{\tau^{\prime}}_{\tau}B_{1}\cdots B_{R}|0\rangle$ 
from the right-hand side of the relation~\eqref{eq:TBBBOmega}.
Then we see how the operator $T^{\tau^{\prime}}_{\tau}$ acts
on the regular Bethe state $|R\rangle$.
By setting $\tau=\tau^{\prime}$ and 
taking the sum on $\tau$, it follows that
the Bethe state $|R\rangle$ is an eigenvector of
the SCP transfer matrix $t(q,\Bar{q}(s);\{p\})$.
\begin{thm}
\label{thm:t-eigenvalue}
Every regular Bethe state $|R\rangle$  
 is an eigenvector of the transfer matrix 
$t(q,\Bar{q}(s);\{p\})$ with $q=(x_{q},y_{q},\mu_{q})$ and
$\Bar{q}(s)=(y_{q},x_{q}\omega^{s},\mu_{q}^{-1})$,
\begin{align}
\label{eq:t-eigenvalue-R_SCP}
&t(q,\Bar{q}(s);\{p\})|R\rangle
 =\sum_{\tau=0}^{N-1}\lambda_{\tau}
  \Big(\prod_{i=1}^{R}
  \mu_{\tau\tau}(z_{i})\Big)|R\rangle
 \nn\\
&=N^{L}\left(\prod_{n=1}^{L}
 \frac{x_{p_{n}}-x_{q}}{x_{p_{n}}^{N}-x_{q}^{N}}
 \frac{y_{p_{n}}-y_{q}}{y_{p_{n}}^{N}-y_{q}^{N}}\right)
 \sum_{\tau=0}^{N-1}
 \left(\prod_{n=1}^{L}
 \frac{t_{p_{n}}^{N}-t_{q}^{N}}{t_{p_{n}}-t_{q}\omega^{\tau}}
 \right)
 \frac{\omega^{\tau(L+R)}F(t_{q})F(t_{q}\omega^{s})}
  {F(t_{q}\omega^{\tau})F(t_{q}\omega^{\tau+1})}
 |R\rangle,
\end{align}
where $F(t)$ is a polynomial in $t$ defined by
$F(t)=\prod_{i=1}^{R}(1-tz_{i})$
and $\{z_{i}|i=1,\ldots,R\}$ is a regular solution
of the Bethe equations~\eqref{eq:Bethe-eq}.
\end{thm}
\begin{proof}
{}By taking the sum of the left-hand side of 
the relation~\eqref{eq:TBBBOmega} with $\tau=\tau^{\prime}$
over $\tau=0,1,\ldots, N-1$ and 
using the result \eqref{eq:t-eigenvalue-0_SCP}, we obtain
\begin{align}
&\sum_{\tau=0}^{N-1}
 \Big(\prod_{i=1}^{R}\mu_{\tau\tau}(z_{i})\Big)
 B_{1}\cdots B_{R}
 T^{\tau}_{\tau}|0\rangle
 =\sum_{\tau=0}^{N-1}\lambda_{\tau}
  \Big(\prod_{i=1}^{R}
  \mu_{\tau\tau}(z_{i})\Big)
  B_{1}\cdots B_{R}|0\rangle.
 \nn
\end{align}
On the other hand, from the right-hand side of 
the relation~\eqref{eq:TBBBOmega}, we have
\begin{align}
&\sum_{\tau=0}^{N-1}
 \Big(\prod_{i=1}^{R}\mu_{\tau\tau}(z_{i})\Big)
 B_{1}\cdots B_{R}
 T^{\tau}_{\tau}|0\rangle
 \nn\\
&=\sum_{\tau=0}^{N-1}\hspace{-3pt}
  \sum_{{\{i_{\ell}\}, \{j_{\ell}\}, \{k_{\ell}\} 
  \atop n_{B}+n_{A}+n_{D}=R}}
  \hspace{-5pt}
  \prod_{p=1}^{n_{A}}
  \frac{\mu_{\tau\tau}(z_{j_{p}})\nu_{\tau+p}(z_{j_{p}})}
  {\mu_{\tau,\tau+p-1}(z_{j_{p}})}
  \prod_{q=1}^{n_{D}}
  \frac{-\mu_{\tau\tau}(z_{k_{q}})\nu_{\tau-q+1}(z_{k_{q}})}
  {\mu_{\tau-q+1,\tau}(z_{k_{q}})}
 \nn\\
&\quad\times
 \Bigg(
 \prod_{i\in\{i_{\ell}\}\atop j\in\{j_{\ell}\}}\!\!
 \omega f_{ji}
 \!\!\prod_{i\in\{i_{\ell}\}\atop k\in\{k_{\ell}\}}\!\!
 \omega f_{ik}
 \!\!\prod_{j\in\{j_{\ell}\}\atop k\in\{k_{\ell}\}}\!\!
 \omega f_{jk}\Bigg)\;
 a_{j_{1}}\cdots a_{j_{n_{A}}}
 d_{k_{1}}\cdots d_{k_{n_{D}}}
 T^{\tau-n_{D}}_{\tau+n_{A}}
 B_{i_{1}}\cdots B_{i_{n_{B}}}|0\rangle
 \nn\\
&=\sum_{m=0}^{R}
  \sum_{{\{i_{\ell}\} \atop n_{B}=m}}
  \Big(\prod_{i\in\Sigma_{R}\setminus\{i_{\ell}\}}
  \hspace{-10pt}a_{i}\Big)
  \sum_{\tau=0}^{N-1}\hspace{-5pt}
  \sum_{{\{j_{\ell}\}, \{k_{\ell}\} \atop n_{A}+n_{D}=R-m}}
  \hspace{-10pt}
  \prod_{p=1}^{n_{A}}
  \frac{\mu_{\tau\tau}(z_{j_{p}})\nu_{\tau+p}(z_{j_{p}})}
  {\mu_{\tau,\tau+p-1}(z_{j_{p}})}
  \prod_{q=1}^{n_{D}}
  \frac{-\mu_{\tau\tau}(z_{k_{q}})\nu_{\tau-q+1}(z_{k_{q}})}
  {\mu_{\tau-q+1,\tau}(z_{k_{q}})}
  \nn\\
&\quad\times
  \omega^{m(R-m)+n_{A}n_{D}}
  \prod_{r=1}^{n_{B}}
  \prod_{p=1}^{n_{A}}
  \prod_{q=1}^{n_{D}}
  f_{j_{p}i_{r}}f_{k_{q}i_{r}}f_{k_{q}j_{p}}\;
  T^{\tau-n_{D}}_{\tau+n_{A}}
  B_{i_{1}}\cdots B_{i_{m}}|0\rangle
 \nn\\
&=\sum_{m=0}^{R}
  \sum_{{\{i_{\ell}\} \atop n_{B}=m}}
  \Big(\prod_{i\in\Sigma_{R}\setminus\{i_{\ell}\}}
  \hspace{-12pt}a_{i}\Big)
  \sum_{\tau=0}^{N-1}\hspace{-5pt}
  \sum_{{\{j_{\ell}\}, \{k_{\ell}\} \atop n_{A}+n_{D}=R-m}}
  \hspace{-10pt}\omega^{m(R-m)+n_{A}n_{D}}
  \nn\\
&\quad\times
  \prod_{p=1}^{n_{A}}
  \frac{\mu_{\tau+n_{D},\tau+n_{D}}(z_{j_{p}})}
  {\mu_{\tau+n_{D},\tau+n_{D}+p-1}(z_{j_{p}})}
  \nu_{\tau+n_{D}+p}(z_{j_{p}})
  \prod_{q=1}^{n_{D}}
  \frac{-\mu_{\tau+n_{D},\tau+n_{D}}(z_{k_{q}})}
  {\mu_{\tau+n_{D}-q+1,\tau+n_{D}}(z_{k_{q}})}
  \nu_{\tau+n_{D}-q+1}(z_{k_{q}})
  \nn\\
&\quad\times
  \prod_{r=1}^{n_{B}}
  \prod_{p=1}^{n_{A}}
  \prod_{q=1}^{n_{D}}
  f_{j_{p}i_{r}}f_{k_{q}i_{r}}f_{k_{q}j_{p}}\;
  T^{\tau}_{\tau+R-m}
  B_{i_{1}}\cdots B_{i_{m}}|0\rangle
 \nn\\
&=\sum_{\tau=0}^{N-1}
  T^{\tau}_{\tau}
  B_{1}\cdots B_{R}|0\rangle
 =t(q,\Bar{q}(s);\{p\})|R\rangle.
\label{eq:proof-t-eigenvalue_SCP}
\end{align}
In the second equality, we have calculated as
\begin{align}
&\Bigg(
 \prod_{i\in\{i_{\ell}\}}
 \prod_{j\in\{j_{\ell}\}}
 \prod_{k\in\{k_{\ell}\}}
 f_{ik}f_{jk}\Bigg)\;
 d_{k_{1}}\cdots d_{k_{n_{D}}}
 =\prod_{k\in\{k_{\ell}\}}
 \Bigg(d_{k}
 \prod_{i\in\{i_{\ell}\}\cup\{j_{\ell}\}}f_{ik}
 \Bigg)\;
 \nn\\
&=\prod_{k\in\{k_{\ell}\}}
 \Bigg(a_{k}
 \prod_{i\in\{i_{\ell}\}\cup\{j_{\ell}\}}f_{ki}
 \Bigg)\;
 =\Bigg(
 \prod_{i_{\ell}\in\{i_{\ell}\}}
 \prod_{j_{\ell}\in\{j_{\ell}\}}
 \prod_{k_{\ell}\in\{k_{\ell}\}}
 f_{ki}f_{kj}\Bigg)\;
 a_{k_{1}}\cdots a_{k_{n_{D}}},
 \nn
\end{align}
where we have employed the identity 
in Lemma~\ref{lm:key-identity_1} which 
is derived from the Bethe equations~\eqref{eq:Bethe-eq}.
The unwanted terms $m\neq R$ 
in~\eqref{eq:proof-t-eigenvalue_SCP} have been canceled out
because of the identity in Lemma~\ref{lm:key-identity_2}.
\end{proof}

\mathversion{bold}
\subsection{The Ising-like spectrum %
consisting of $2^{r}$ eigenvalues}
\mathversion{normal}

 From the expression of eigenvalues of the SCP 
transfer matrix $t(q,\Bar{q}(s);\{p\})$ 
we can derive eigenvalues of the diagonal-to-diagonal 
transfer matrices $T_{\text{D}}(x_{q},y_{q})$ 
and $\Hat{T}_{\text{D}}(x_{q},y_{q})$.
 {} From the discussion similar to \cite{Baxter_93JSP},
the set of eigenvalues in the invariant subspace containing 
a given regular Bethe state 
$|R\rangle$ are given in the following forms:
\begin{align}
\label{eq:T-form}
&\Lambda(x_{q},y_{q})=N^{\frac{L}{2}}
 \left(\prod_{n=1}^{L}
 \frac{x_{p_{n}}-x_{q}}{x_{p_{n}}^{N}-x_{q}^{N}}\right)
 x_{q}^{P_{a}}y_{q}^{P_{b}}\mu_{q}^{-NP_{c}}
 F(t_{q})G(\mu_{q}^{-N}),
 \nn\\
&\Hat{\Lambda}(x_{q},y_{q})=N^{\frac{L}{2}}
 \left(\prod_{n=1}^{L}
 \frac{y_{p_{n}}-x_{q}}{y_{p_{n}}^{N}-x_{q}^{N}}\right)
 x_{q}^{P_{a}}y_{q}^{P_{b}}\mu_{q}^{-NP_{c}}
 F(t_{q})\Hat{G}(\mu_{q}^{-N}).
\end{align}
Here $P_{a}$ and $P_{b}$ are integers satisfying 
$P_{a}+P_{b}\equiv -L-R\mod N$,
and we recall $F(t_{q})=\prod_{i=1}^{R}(1-t_{q}z_{i})$. 
$G(\mu_{q}^{N})$ and $\Hat{G}(\mu_{q}^{N})$ 
are polynomials in $\mu_{q}^{N}$
satisfying $G(\mu_{q}^{N})
=\mathrm{const.}\Hat{G}(\mu_{q}^{N})$.

{}From the relation $t(q,\Bar{q}(s);\{p\})
=T_{\text{D}}(x_{q},y_{q})
\Hat{T}_{\text{D}}(y_{q},x_{q}\omega^{s})$, 
the product $G(\mu_{q}^{-N})\Hat{G}(\mu_{q}^{N})$
is given by 
\begin{align}
\label{eq:SCP-poly}
&G(\mu_{q}^{-N})\Hat{G}(\mu_{q}^{N})
=P_{\text{SCP}}(t_{q}^{N})
\Define\sum_{\tau=0}^{N-1}
 \left(\prod_{n=1}^{L}
 \frac{t_{p_{n}}^{N}-t_{q}^{N}}{t_{p_{n}}-t_{q}\omega^{\tau}}
 \right)
 \frac{\omega^{\tau(L+R)}}
  {F(t_{q}\omega^{\tau})F(t_{q}\omega^{\tau+1})}.
\end{align}
Here $P_{\text{SCP}}(t_{q}^{N})$ 
is a polynomial in $t_{q}^{N}$ of degree at most 
$\lfloor\frac{L(N-1)-2R}{N}\rfloor$;
the Bethe equations~\eqref{eq:Bethe-eq} correspond to
the pole-free condition.
We call the polynomial $P_{\text{SCP}}(\zeta)$ the SCP polynomial. 
We remark that, in our result,
only the case $P_{b}=0$ appears.
The relation $k^{2}t_{q}^{N}
=1-k^{\prime}(\mu_{q}^{N}+\mu_{q}^{-N})+k^{\prime 2}$
tells us that the polynomial $P_{\text{SCP}}(t_{q}^{N})$ 
is regarded as a Laurent polynomial in $\mu_{q}^{N}$ 
of degree $r=\mathrm{deg}P_{\text{SCP}}(\zeta)$
whose zeros occur in reciprocal pairs.
Then, by denoting the $2r$ zeros by $\{w_{i}^{\pm 1}\}$,
we have $2^{r}$ solutions for $G(\mu_{q}^{N})$ and 
$\Hat{G}(\mu_{q}^{N})$ in the forms
\begin{align}
\label{eq:GG_SCP}
 G(\mu_{q}^{N}), \Hat{G}(\mu_{q}^{N})
 =\text{const.}\prod_{i=1}^{r}
 (\mu_{q}^{N}-w_{i}^{\epsilon_{i}}),
\end{align}
where $\epsilon_{i}=1$ or $-1$
is independently chosen for the index $i$.

The $2^{r}$ solutions for $G(\mu_{q}^{N})$ and  
$\Hat{G}(\mu_{q}^{N})$ are similar to 
the $2^{r}$ eigenvalues of the Ising-like 
form~\cite{Onsager_44PR,Davies_90JPA}. 
We thus call the set of $2^{r}$ eigenvalues of 
the diagonal-to-diagonal transfer matrices associated 
with a regular Bethe state 
the Ising-like spectrum associated with the regular Bethe state. 
In fact, in the homogeneous case of $p_{1}=\cdots=p_{L}$, 
it follows from the Onsager-algebra structure of the SCP model 
that each eigenvalue is non-degenerate, that is, 
the multiplicity of the eigenvalue specified by 
a set of $\{\epsilon_{i}\}$ is given by one.

For the homogeneous case, the Ising-like spectrum 
of the diagonal-to-diagonal transfer matrix was shown  
by applying the functional relations among 
the transfer matrices~\cite{%
Albertini-McCoy-Perk-Tang_89NPB,Albertini-McCoy-Perk_89ASPM,%
Tarasov_90PLA,Baxter_88PLA,Baxter_89JSP,Baxter_93JSP}.
There are three types of the functional relations~\cite{Baxter_88PLA,Baxter_89JSP,Baxter_93JSP}: 
the first relation is based on the fact that 
the transfer matrix of the SCP model is exactly a $Q$-operator for 
the $\tau_{2}$-model~\cite{Bazhanov-Stroganov_90NPB,Roan_07JSM}, 
and it gives eigenvalues of the transfer matrix of the $\tau_{2}$-model.
The second relation is interpreted as a 
$T$-system~\cite{Kirillov-Reshetikhin_85JSM,%
Klumper-Pearce_92PhysicaA}, 
which recursively generates the eigenvalues of 
the transfer matrices in the fusion hierarchy. 
The third relation leads to the eigenvalues of 
the product of the diagonal-to-diagonal transfer matrices 
of the SCP model with a constraint on the spectral parameters.
The algebraic Bethe ansatz of the $\tau_{2}$-model 
given in the previous section 
plays the same role as the first functional 
relation~\cite{Tarasov_90PLA}.
The algebraic approach formulated in this section 
plays a similar role as the second and third functional relations. 

\mathversion{bold}
\subsection{Proof of Proposition~\ref{prop:TBBBOmega-relation}}
\mathversion{normal}

The subsection is devoted to
a proof of Proposition~\ref{prop:TBBBOmega-relation}.
Our strategy is to derive a recursion relation
for the coefficients $c^{\tau^{\prime}\tau}_{n}
(\{i_{\ell}\};\{j_{\ell}\};\{k_{\ell}\})$
in the relation~\eqref{eq:TBBBOmega}.

\begin{lm}
\label{prop:4rels-YB}
The Yang-Baxter relation \eqref{eq:LTT=TTL} is
equivalent to the following relations:
\begin{align}
&\alpha_{\tau^{\prime}}(z)A(z)T_{\tau}^{\tau^{\prime}}
 +\beta_{\tau^{\prime}}(z)C(z)T_{\tau}^{\tau^{\prime}-1} 
 =\alpha_{\tau}(z)T_{\tau}^{\tau^{\prime}}A(z)
 +\gamma_{\tau}(z)T_{\tau-1}^{\tau^{\prime}}B(z), \nn\\
&\alpha_{\tau^{\prime}}(z)B(z)T_{\tau}^{\tau^{\prime}}
 +\beta_{\tau^{\prime}}(z)D(z)T_{\tau}^{\tau^{\prime}-1} 
 =\beta_{\tau+1}(z)T_{\tau+1}^{\tau^{\prime}}A(z)
 +\delta_{\tau}(z)T_{\tau}^{\tau^{\prime}}B(z), \nn\\
&\gamma_{\tau^{\prime}+1}(z)A(z)T_{\tau}^{\tau^{\prime}+1}
 +\delta_{\tau^{\prime}}(z)C(z)T_{\tau}^{\tau^{\prime}}
 =\alpha_{\tau}(z)T_{\tau}^{\tau^{\prime}}C(z)
 +\gamma_{\tau}(z)T_{\tau-1}^{\tau^{\prime}}D(z), \nn\\
&\gamma_{\tau^{\prime}+1}(z)B(z)T_{\tau}^{\tau^{\prime}+1}
 +\delta_{\tau^{\prime}}(z)D(z)T_{\tau}^{\tau^{\prime}}
 =\beta_{\tau+1}(z)T_{\tau+1}^{\tau^{\prime}}C(z)
 +\delta_{\tau}(z)T_{\tau}^{\tau^{\prime}}D(z),
\end{align}
where
\begin{align}
&\alpha_{\tau}(z)
 =-y_{q_{1}}y_{q_{2}}z
  +\mu_{q_{1}}\mu_{q_{2}}\omega^{\tau}, \qquad
 \beta_{\tau}(z)
 =-z(y_{q_{1}}-x_{q_{2}}\mu_{q_{1}}\mu_{q_{2}}\omega^{\tau}), \nn\\
&\gamma_{\tau}(z)
 =y_{q_{2}}-x_{p_{1}}\mu_{q_{1}}\mu_{q_{2}}\omega^{\tau}, \qquad
 \delta_{\tau}(z)
 =1-x_{q_{1}}x_{q_{2}}\mu_{q_{1}}\mu_{q_{2}}z\omega^{\tau+1}. \nn
\end{align}
Here we have omitted the dependence of the 
spectral parameters $q_{1}$ and $q_{2}$ in the coefficients 
$\alpha_{\tau}(z), \beta_{\tau}(z), \gamma_{\tau}(z), 
\delta_{\tau}(z)$ and the operator $T_{\tau}^{\tau^{\prime}}$.
\end{lm}

\begin{lm}
For the operators $T^{\tau^{\prime}}_{\tau}$, we have
\begin{align}
\label{eq:TB-relation}
&\mu_{\tau^{\prime}\tau}(z)B(z)T^{\tau^{\prime}}_{\tau}
 \nn\\
&=T^{\tau^{\prime}}_{\tau}B(z)
 +\nu_{\tau+1}(z)T^{\tau^{\prime}}_{\tau+1}A(z)
 -\nu_{\tau^{\prime}}(z)T^{\tau^{\prime}-1}_{\tau}D(z)
 -\nu_{\tau^{\prime}}(z)\nu_{\tau+1}(z)
  T^{\tau^{\prime}-1}_{\tau+1}C(z) . 
\end{align}
Here $\mu_{\tau^{\prime}\tau}(z)$ and 
$\nu_{\tau}(z)$ coincide with those defined in \eqref{eq:mu-nu}, 
respectively, by setting $q_{1}=(x_{q},y_{q},\mu_{q})$ and 
$q_{2}=(y_{q},x_{q}\omega^{s},\mu_{q}^{-1})$. 
\end{lm}
\begin{proof}
{}From the second and fourth relations 
in Lemma~\ref{prop:4rels-YB}, we have
\begin{align}
T_{\tau}^{\tau^{\prime}}B(z)
&=\left(\frac{\alpha_{\tau^{\prime}}(z)}{\delta_{\tau}(z)}
  -\frac{\beta_{\tau^{\prime}}(z)}{\delta_{\tau}(z)}
  \frac{\gamma_{\tau^{\prime}}(z)}{\delta_{\tau^{\prime}-1}(z)}
  \right)
  B(z)T_{\tau}^{\tau^{\prime}}
 -\frac{\beta_{\tau+1}(z)}{\delta_{\tau}(z)}
  T_{\tau+1}^{\tau^{\prime}}A(z)
 \nn\\
&\quad
 +\frac{\beta_{\tau^{\prime}}(z)}{\delta_{\tau^{\prime}-1}(z)}
  T_{\tau}^{\tau^{\prime}-1}D(z)
 +\frac{\beta_{\tau^{\prime}}(z)}{\delta_{\tau^{\prime}-1}(z)}
  \frac{\beta_{\tau+1}(z)}{\delta_{\tau}(z)}
  T_{\tau+1}^{\tau^{\prime}-1}C(z).
 \nn
\end{align}
By setting $q_{1}=(x_{q},y_{q},\mu_{q})$ and
$q_{2}=(y_{q},x_{q}\omega^{s},\mu_{q}^{-1})$, 
we prove the relation.
\end{proof}

\begin{lm}
Let $I=\{i_{\ell}\}$, $J=\{j_{\ell}\}$ and $K=\{k_{\ell}\}$ 
be such disjoint subsets of 
the set $\Sigma_{n}=\{1,2,\ldots,n\}$ that
$\sharp I=n_{B}$, $\sharp J=n_{A}$, $\sharp K=n_{D}$ 
and $n_{B}+n_{A}+n_{D}=n$.
The coefficients $c_{n}(I;J;K)=c^{\tau^{\prime}\tau}_{n}(I;J;K)$ 
in the relation~\eqref{eq:TBBBOmega}
satisfy the following recursion relation on $n$:
\begin{align}
&c_{n}
 (I;J;K)
 \nn\\
&=c_{n-1}
  (I\!\setminus\!\{n\};J;K)
  \frac{1}{\mu_{\tau^{\prime}\!-n_{D},\tau+n_{A}}(z_{n})}
 \nn\\
&+c_{n-1}
  (I;J\!\setminus\!\{n\};K)
  \frac{\nu_{\tau+n_{A}}(z_{n})}
  {\mu_{\tau^{\prime}\!-n_{D},\tau+n_{A}-1}(z_{n})}
  \prod_{i\in I}\omega f_{ni}
 \nn\\
&-\sum_{j\in J}
  c_{n-1}
  (I\!\cup\!\{j\}\!\setminus\!\{n\};J\!\setminus\!\{j\};K)
  \frac{\nu_{\tau+n_{A}}(z_{n})}
  {\mu_{\tau^{\prime}\!-n_{D},\tau+n_{A}-1}(z_{n})}
  \omega \Big(\prod_{i\in I\setminus\{n\}}
  \omega f_{ji}\Big)
  g_{nj}
 \nn\\
&-c_{n-1}(I;J;K\!\setminus\!\{n\})
  \frac{\nu_{\tau^{\prime}-n_{D}+1}(z_{n})}
  {\mu_{\tau^{\prime}\!-n_{D}+1,\tau+n_{A}}(z_{n})}
  \prod_{i\in I}\omega f_{in}
 \nn\\
&-\sum_{k\in K}
  c_{n-1}
  (I\!\cup\!\{k\}\!\setminus\!\{n\};
   J;K\!\setminus\!\{k\})
  \frac{\nu_{\tau^{\prime}-n_{D}+1}(z_{n})}
  {\mu_{\tau^{\prime}\!-n_{D}+1,\tau+n_{A}}(z_{n})}
  \omega \Big(\prod_{i\in I\setminus\{n\}}
  \omega f_{ik}\Big)
  g_{nk}
 \nn\\
&-\sum_{k\in K}
  c_{n-1}
  (I\!\cup\!\{k\};
   J\!\setminus\!\{n\};
   K\!\setminus\!\{k\})
 \frac{\nu_{\tau^{\prime}\!-n_{D}+1}(z_{n})\nu_{\tau+n_{A}}(z_{n})}
  {\mu_{\tau^{\prime}\!-n_{D}+1,\tau+n_{A}-1}(z_{n})}
 \omega \Big(\prod_{i\in I}
 \omega f_{ni}\omega f_{ik}\Big)
 g_{nk}
 \nn\\
&+\sum_{j\in J}
  c_{n-1}
  (I\!\cup\!\{j\};
   J\!\setminus\!\{j\};
   K\!\setminus\!\{n\})
 \frac{\nu_{\tau^{\prime}\!-n_{D}+1}(z_{n})\nu_{\tau+n_{A}}(z_{n})}
  {\mu_{\tau^{\prime}\!-n_{D}+1,\tau+n_{A}-1}(z_{n})}
 \omega \Big(\prod_{i\in I}
 \omega f_{ji}\omega f_{in}\Big)
 g_{nj}
 \nn\\
&+\sum_{j\in J\atop k\in K}
  c_{n-1}
  (I\!\cup\!\{j\}\!\cup\!\{k\}\!\setminus\!\{n\};
   J\!\setminus\!\{j\};
   K\!\setminus\!\{k\})
 \nn\\
&\qquad\quad\times
 \frac{\nu_{\tau^{\prime}\!-n_{D}+1}(z_{n})\nu_{\tau+n_{A}}(z_{n})}
  {\mu_{\tau^{\prime}\!-n_{D}+1,\tau+n_{A}-1}(z_{n})}
 \omega f_{jk}
 \Big(\prod_{i\in I\setminus\{n\}}
 \omega f_{ji}\omega f_{ik}\Big)g_{nj}g_{nk}.
 \nn
\end{align}
Here, if the set $S_{1}$ is not a subset of $S$,
we set $c_{n}(S\setminus S_{1};\cdot;\cdot)=
c_{n}(\cdot;S\setminus S_{1};\cdot)=
c_{n}(\cdot;\cdot;S\setminus S_{1})=0$.
\end{lm}
\begin{proof}
We apply the operator $B_{n}=B(z_{n})$ 
to both sides of the equation~\eqref{eq:TBBBOmega} 
with $n-1$ in place of $n$.
Let $\Tilde{I}=\{\Tilde{i}_{\ell}\}$, 
$\Tilde{J}=\{\Tilde{j}_{\ell}\}$ 
and $\Tilde{K}=\{\Tilde{k}_{\ell}\}$ 
be such disjoint subsets of 
the set $\Sigma_{n-1}$ that
$\sharp \Tilde{I}=m_{B}$, 
$\sharp \Tilde{J}=m_{A}$, 
$\sharp \Tilde{K}=m_{D}$ 
and $m_{B}+m_{A}+m_{D}=n-1$.
By using the relation \eqref{eq:TB-relation},
we have
\begin{align}
&B_{1}\cdots B_{n-1}B_{n}T^{\tau^{\prime}}_{\tau}
 |0\rangle
 \nn\\
&=\!\sum_{\Tilde{I}, \Tilde{J}, \Tilde{K}}\!
  c^{\tau^{\prime}\tau}_{n-1}
  (\Tilde{I};\Tilde{J};\Tilde{K})
  B_{n}
  T^{\tau^{\prime}-m_{D}}_{\tau+m_{A}}
  B_{\Tilde{i}_{1}}\cdots B_{\Tilde{i}_{m_{B}}}
  A_{\Tilde{j}_{1}}\cdots A_{\Tilde{j}_{m_{A}}}
  D_{\Tilde{k}_{1}}\cdots D_{\Tilde{k}_{m_{D}}}
  |0\rangle
 \nn\\
&=\!\sum_{\Tilde{I}, \Tilde{J}, \Tilde{K}}\!
  c^{\tau^{\prime}\tau}_{n-1}(\Tilde{I};\Tilde{J};\Tilde{K})
  \Big(
  \frac{1}{\mu_{\tau^{\prime}-m_{D},\tau+m_{A}}(z_{n})}
  T^{\tau^{\prime}-m_{D}}_{\tau+m_{A}}B_{n}
  +\frac{\nu_{\tau+m_{A}+1}(z_{n})}
   {\mu_{\tau^{\prime}-m_{D},\tau+m_{A}}(z_{n})}
   T^{\tau^{\prime}-m_{D}}_{\tau+m_{A}+1}A_{n}
 \nn\\
&\qquad
  -\frac{\nu_{\tau^{\prime}-m_{D}}(z_{n})}
   {\mu_{\tau^{\prime}-m_{D},\tau+m_{A}}(z_{n})}
   T^{\tau^{\prime}-m_{D}-1}_{\tau+m_{A}}D_{n}
  -\frac{\nu_{\tau^{\prime}-m_{D}}(z_{n})
   \nu_{\tau+m_{A}+1}(z_{n})}
   {\mu_{\tau^{\prime}-m_{D},\tau+m_{A}}(z_{n})}
   T^{\tau^{\prime}-m_{D}-1}_{\tau+m_{A}+1}C_{n}
  \Big)
 \nn\\
&\qquad\times
  B_{\Tilde{i}_{1}}\cdots B_{\Tilde{i}_{m_{B}}}
  A_{\Tilde{j}_{1}}\cdots A_{\Tilde{j}_{m_{A}}}
  D_{\Tilde{k}_{1}}\cdots D_{\Tilde{k}_{m_{D}}}
  |0\rangle.
\nn
\end{align}
By arranging the operators $A(z)$, $B(z)$, $C(z)$ and $D(z)$
in the order $BADC$ with the relations in Lemma~\ref{lm:ABCD} 
and by rewriting the terms
in the form of the right-hand side of \eqref{eq:TBBBOmega},
we obtain the recursion relation.
\end{proof}

We now prove Proposition~\ref{prop:TBBBOmega-relation}.
{}From the symmetry of the relation \eqref{eq:TBBBOmega},
it is enough to solve the recursion relation in the case 
$i_{1}<\cdots<i_{n_{B}}<j_{1}<\cdots<j_{n_{A}}
<k_{1}<\cdots<k_{n_{D}}$.
First we consider the case $n_{B}=n$, 
that is, $i_{\ell}=\ell$ for $\ell=1,2,\ldots,n$
and $J=K=\phi$.
The recursion relation is reduced to
\begin{align}
&c_{n}(I;\phi;\phi)
 =c_{n-1}(I\!\setminus\!\{n\};\phi;\phi)
  \frac{1}{\mu_{\tau^{\prime}\tau}(z_{n})}.
 \nn
\end{align}
{}From the initial condition $c_{0}(\phi;\phi;\phi)=1$, 
the recursion relation is solved as
\[
 c_{n}(I;\phi;\phi)
 =\prod_{i\in I}\frac{1}{\mu_{\tau^{\prime}\tau}(z_{i})},
\]
which is consistent with the form \eqref{eq:TBBBOmega}.
Second we consider the case $n_{B}+n_{A}=n$, 
that is, $n\in\{j_{\ell}\}$ and $K=\phi$.
The recursion relation is reduced to
\begin{align}
c_{n}(I;J;\phi)
&=c_{n-1}(I;J\!\setminus\!\{n\};\phi)
  \frac{\nu_{\tau+n_{A}}(z_{n})}
  {\mu_{\tau^{\prime},\tau+n_{A}-1}(z_{n})}
  \prod_{i\in I}\omega f_{ni}.
 \nn
\end{align}
By using the result in the case $n_{B}=n$, 
we obtain
\begin{align}
c_{n}(I;J;\phi)
&=\prod_{i\in I}
  \frac{1}{\mu_{\tau^{\prime}\tau}(z_{i})}
  \prod_{p=1}^{n_{A}}
  \frac{\nu_{\tau+p}(z_{j_{p}})}
  {\mu_{\tau^{\prime},\tau+p-1}(z_{j_{p}})}
 \prod_{i\in I\atop j\in J}\omega f_{ji},
 \nn
\end{align}
which is also consistent with the form \eqref{eq:TBBBOmega}.
Third we consider the case $n\in\{k_{\ell}\}$. 
The recursion relation is reduced to
\begin{align}
\label{eq:proof-cn_1}
&c_{n}(I;J;K)=
 -c_{n-1}
  (I;J;K\!\setminus\!\{n\})
  \frac{\nu_{\tau^{\prime}-n_{D}+1}(z_{n})}
  {\mu_{\tau^{\prime}-n_{D}+1,\tau+n_{A}}(z_{n})}
  \prod_{i\in I}\omega f_{in}
 \nn\\
&+\sum_{j\in J}
  c_{n-1}(I\!\cup\!\{j\};J\!\setminus\!\{j\};K\!\setminus\!\{n\})
 \frac{\nu_{\tau^{\prime}-n_{D}+1}(z_{n})\nu_{\tau+n_{A}}(z_{n})}
  {\mu_{\tau^{\prime}-n_{D}+1,\tau+n_{A}-1}(z_{n})}
 \omega
 \Big(\prod_{i\in I}
 \omega f_{ji}\,\omega f_{in}\Big)g_{nj}.
\end{align}
Note that, in the case, the coefficients 
$c_{n-1}(\{i_{\ell}^{\prime}\};
\{j_{\ell}^{\prime}\};\{k_{\ell}^{\prime}\})$
with general sets $\{i_{\ell}^{\prime}\}$,
$\{j_{\ell}^{\prime}\}$ and $\{k_{\ell}^{\prime}\}$,
which are not necessarily in the order
$i_{1}^{\prime}<\cdots<i_{n_{B}}^{\prime}
<j_{1}^{\prime}<\cdots<j_{n_{A}}^{\prime}
<k_{1}^{\prime}<\cdots<k_{n_{D}}^{\prime}$, appear.
Assume that the coefficients $c_{n-1}(I;J;K\setminus\{n\})$
and $c_{n-1}(I\cup\{j\};J\!\setminus\!\{j\};
K\!\setminus\!\{n\})$ in \eqref{eq:proof-cn_1}
are given in the form \eqref{eq:TBBBOmega}.
Substituting the form of the coefficient $c_{n-1}
(\{i_{\ell}\};\{j_{\ell}\};\{k_{\ell}\}\setminus\{n\})$
and $k_{n_{D}}=n$ into the first term of 
\eqref{eq:proof-cn_1}, we obtain
\begin{align}
\label{eq:proof-cn_2}
&-c_{n-1}
  (\{i_{\ell}\};\{j_{\ell}\};\{k_{\ell}\}\!\setminus\!\{n\})
  \frac{\nu_{\tau^{\prime}-n_{D}+1}(z_{n})}
  {\mu_{\tau^{\prime}-n_{D}+1,\tau+n_{A}}(z_{n})}
  \prod_{r=1}^{n_{B}}\omega f_{i_{r}n}
 \nn\\
&=-(-)^{n_{D}-1}\prod_{r=1}^{n_{B}}
  \frac{1}{\mu_{\tau^{\prime}\tau}(z_{i_{r}})}
  \prod_{p=1}^{n_{A}}
  \frac{\nu_{\tau+p}(z_{j_{p}})}
  {\mu_{\tau^{\prime},\tau+p-1}(z_{j_{p}})}
  \prod_{q=1}^{n_{D}-1}
  \frac{\nu_{\tau^{\prime}-q+1}(z_{k_{q}})}
  {\mu_{\tau^{\prime}-q+1,\tau}(z_{k_{q}})}
  \frac{\nu_{\tau^{\prime}-n_{D}+1}(z_{n})}
  {\mu_{\tau^{\prime}-n_{D}+1,\tau+n_{A}}(z_{n})}
 \nn\\
&\quad\times
 \prod_{i\in\{i_{\ell}\}\atop j\in\{j_{\ell}\}}\!\!
 \omega f_{ji}\!\!\!
 \prod_{i\in\{i_{\ell}\}\atop k\in\{k_{\ell}\}\setminus\{n\}}
 \!\!\!\omega f_{ik}\!\!\!
 \prod_{j\in\{j_{\ell}\}\atop k\in\{k_{\ell}\}\setminus\{n\}}
 \!\!\!\omega f_{jk}
 \prod_{r=1}^{n_{B}}\omega f_{i_{r}n}
 \nn\\
&=c_{n}(\{i_{\ell}\};\{j_{\ell}\};\{k_{\ell}\})
 \frac{\mu_{\tau^{\prime}-n_{D}+1,\tau}(z_{n})}
 {\mu_{\tau^{\prime}-n_{D}+1,\tau+n_{A}}(z_{n})}
 \prod_{j\in\{j_{\ell}\}}\frac{1}{\omega f_{jn}}.
\end{align}
In a similar way, substituting the forms of 
the coefficients $c_{n-1}(\{i_{\ell}\}\cup\{j_{p}\};
\{j_{\ell}\}\setminus\{j_{p}\};\{k_{\ell}\}\setminus\{n\})$
and $k_{n_{D}}=n$ into the second term of 
\eqref{eq:proof-cn_1}, we obtain
\begin{align}
\label{eq:proof-cn_3}
&\sum_{p=1 \atop j_{p}\neq n}^{n_{A}}
  c_{n-1}
  (\{i_{\ell}\}\!\cup\!\{j_{p}\};
   \{j_{\ell}\}\!\setminus\!\{j_{p}\};
   \{k_{\ell}\}\!\setminus\!\{n\})
 \frac{\nu_{\tau^{\prime}-n_{D}+1}(z_{n})\nu_{\tau+n_{A}}(z_{n})}
 {\mu_{\tau^{\prime}-n_{D}+1,\tau+n_{A}-1}(z_{n})}
 \omega\Big(\prod_{r=1}^{n_{B}}
 \omega f_{j_{p}i_{r}}\omega f_{i_{r}n}\Big)g_{nj_{p}}
 \nn\\
&=\sum_{p=1 \atop j_{p}\neq n}^{n_{A}}
  (-)^{n_{D}-1}\prod_{r=1}^{n_{B}}
  \frac{1}{\mu_{\tau^{\prime}\tau}(z_{i_{r}})}
  \frac{1}{\mu_{\tau^{\prime}\tau}(z_{j_{p}})}
  \prod_{p^{\prime}=1}^{p-1}
  \frac{\nu_{\tau+p^{\prime}}(z_{j_{p^{\prime}}})}
  {\mu_{\tau^{\prime},\tau+p^{\prime}-1}(z_{j_{p^{\prime}}})}\!
  \prod_{p^{\prime}=p+1}^{n_{A}}\!
  \frac{\nu_{\tau+p^{\prime}-1}(z_{j_{p^{\prime}}})}
  {\mu_{\tau^{\prime},\tau+p^{\prime}-2}(z_{j_{p^{\prime}}})}\!\!
  \prod_{q=1}^{n_{D}-1}\!\!
  \frac{\nu_{\tau^{\prime}-q+1}(z_{k_{q}})}
  {\mu_{\tau^{\prime}-q+1,\tau}(z_{k_{q}})}
 \nn\\
&\quad\times
 \frac{\nu_{\tau^{\prime}-n_{D}+1}(z_{n})\nu_{\tau+n_{A}}(z_{n})}
  {\mu_{\tau^{\prime}-n_{D}+1,\tau+n_{A}-1}(z_{n})}
 \omega
 \prod_{i\in\{i_{\ell}\}\cup\{j_{p}\}
 \atop j\in\{j_{\ell}\}\setminus\{j_{p}\}}
 \!\!\!\omega f_{ji}\!\!\!
 \prod_{i\in\{i_{\ell}\}\cup\{j_{p}\}
 \atop k\in\{k_{\ell}\}\setminus\{n\}}
 \!\!\!\omega f_{ik}\!\!\!
 \prod_{j\in\{j_{\ell}\}\setminus\{j_{p}\}
 \atop k\in\{k_{\ell}\}\setminus\{n\}}
 \!\!\!\omega f_{jk}
 \Big(\prod_{r=1}^{n_{B}}
 \omega f_{j_{p}i_{r}}\omega f_{i_{r}n}\Big)g_{nj_{p}}
 \nn\\
&=-c_{n}(\{i_{\ell}\};\{j_{\ell}\};\{k_{\ell}\})
 \sum_{p=1}^{n_{A}}
 \frac{\mu_{\tau^{\prime},\tau+n_{A}-1}(z_{j_{p}})}
 {\mu_{\tau^{\prime}\tau}(z_{j_{p}})}
 \frac{\mu_{\tau^{\prime}-n_{D}+1,\tau}(z_{n})}
 {\mu_{\tau^{\prime}-n_{D}+1,\tau+n_{A}-1}(z_{n})}
 \frac{\nu_{\tau+n_{A}}(z_{n})}
 {\nu_{\tau+n_{A}}(z_{j_{p}})}
 \frac{g_{nj_{p}}}{f_{j_{p}n}}
 \prod_{j\in\{j_{\ell}\}\setminus\{j_{p}\}}
 \frac{f_{jj_{p}}}{f_{jn}}.
\end{align}
Hence, by combining \eqref{eq:proof-cn_2}
and \eqref{eq:proof-cn_3}, the coefficient $c_{n}
(\{i_{\ell}\};\{j_{\ell}\};\{k_{\ell}\})$ is shown to be
in the form \eqref{eq:TBBBOmega} if
the following relation holds:
\begin{align}
&\frac{\mu_{\tau^{\prime}-n_{D}+1,\tau}(z_{n})}
 {\mu_{\tau^{\prime}-n_{D}+1,\tau+n_{A}}(z_{n})}
 \prod_{p=1}^{n_{A}}\frac{1}{\omega f_{j_{p}n}}
 \nn\\
&-\sum_{p=1}^{n_{A}}
 \frac{\mu_{\tau^{\prime},\tau+n_{A}-1}(z_{j_{p}})}
 {\mu_{\tau^{\prime}\tau}(z_{j_{p}})}
 \frac{\mu_{\tau^{\prime}-n_{D}+1,\tau}(z_{n})}
 {\mu_{\tau^{\prime}-n_{D}+1,\tau+n_{A}-1}(z_{n})}
 \frac{\nu_{\tau+n_{A}}(z_{n})}
 {\nu_{\tau+n_{A}}(z_{j_{p}})}
 \frac{g_{nj_{p}}}{f_{j_{p}n}}
 \prod_{j\in\{j_{\ell}\}\setminus\{j_{p}\}}
 \frac{f_{jj_{p}}}{f_{jn}}=1,
 \nn
\end{align}
which is the identity in Lemma~\ref{lm:key-identity_1}.

\mathversion{bold}
\section{The $\mathfrak{sl}_{2}$ loop algebra symmetry 
of the $\tau_{2}$-model and degenerate eigenspaces}
\mathversion{normal}
\label{sec:degenerate}

\mathversion{bold}
\subsection{Gauge transformations on the $L$-operator}
\mathversion{normal}

We now introduce another $L$-operator in order to show the 
$\mathfrak{sl}_{2}$ loop algebra symmetry
of the $\tau_{2}$-model. 
The degenerate eigenspace of the transfer matrix 
constructed from the new $L$-operator is identical to 
the degenerate eigenspace of the $\tau_{2}$-model which 
we have introduced in Section~\ref{sec:def_tau-2-model}. 

Let us introduce the $L$-operator $\Tilde{\mathcal{L}}_{i}(z)
\in\End(\mathbb{C}^{2}\otimes(\mathbb{C}^{N})^{\otimes L}),
(i=1,2,\ldots,L)$ given by
\begin{align}
 \Tilde{\mathcal{L}}_{i}(z)
&=\begin{pmatrix}
   q^{-\frac{1}{2}}\big(z(k^{\frac{1}{2}})_{i}
   -z^{-1}(k^{-\frac{1}{2}})_{i}\big)
  &(q-q^{-1})(f)_{i} \\
   (q-q^{-1})(e)_{i} 
  &q^{\frac{1}{2}}\big(z(k^{-\frac{1}{2}})_{i}
   -z^{-1}(k^{\frac{1}{2}})_{i}\big) \\
  \end{pmatrix}.
\end{align}
Here $q$ is not a rapidity on the Fermat 
curve~\eqref{eq:Fermat} but a generic parameter,
and $\{(k)_{i},(e)_{i},(f)_{i}\}$ is the $N$-dimensional
representation of the quantum algebra $U_{q}(\mathfrak{sl}_{2})$
non-trivially acting only on the $i$th component of 
the quantum space $(\mathbb{C}^{N})^{\otimes L}$ as
\[
 kv_{\sigma}=\varepsilon q^{N-1-2\sigma}v_{\sigma},\qquad
 ev_{\sigma}=\varepsilon \alpha [N-\sigma]v_{\sigma-1},\qquad
 fv_{\sigma}=\alpha^{-1}[\sigma+1]v_{\sigma+1},
\]
with $\alpha\neq 0$ 
and $[n]=\frac{q^{n}-q^{-n}}{q-q^{-1}}$.
We set $\varepsilon=1$ for odd $N$
and $\varepsilon=-1$ for even $N$.
One sees that the $L$-operator $\Tilde{\mathcal{L}}_{i}(z)$
is nothing but that of an XXZ spin chain
with $N$-state local spins and a twist parameter.
The $L$-operator $\Tilde{\mathcal{L}}_{i}(z)$
satisfies the Yang-Baxter relation~\eqref{eq:RLL=LLR} with 
the $R$-matrix of the six-vertex model given by
\begin{align}
\label{eq:R-matrix_6v}
 R_{\mathrm{6v}}(z)=
 \begin{pmatrix}
  1-z^{2}q^{2} & 0 & 0 & 0 \\
  0 & (1-z^{2})q & z(1-q^{2}) & 0 \\
  0 & z(1-q^{2}) & (1-z^{2})q & 0 \\
  0 & 0 & 0 & 1-z^{2}q^{2} \\
 \end{pmatrix}
 .
\end{align}
We introduce the monodromy matrix 
$\Tilde{\mathcal{T}}(z;\{p\})\in\End\big(\mathbb{C}^{2}
\otimes(\mathbb{C}^{N})^{\otimes L}\big)$
and the transfer matrix $\Tilde{\tau}(z)=
\Tilde{\tau}(z;\{p\})\in\End\big(
(\mathbb{C}^{N})^{\otimes L}\big)$ as
\begin{align}
&\Tilde{\mathcal{T}}(z;\{p\})=
 \prod_{i=1}^{L}
 \Tilde{\mathcal{L}}_{i}
 (t_{p_{i}}^{\frac{1}{2}}zq^{\frac{1}{2}})
 \definE
 \begin{pmatrix}
  \Tilde{A}(z) & \Tilde{B}(z) \\
  \Tilde{C}(z) & \Tilde{D}(z) 
 \end{pmatrix}
 ,\quad
 \Tilde{\tau}(z;\{p\})=\tr_{\mathbb{C}^{2}}
 \big(\Tilde{\mathcal{T}}(z;\{p\})\big).
\end{align}
In a way similar to Section~\ref{sec:def_tau-2-model},
we apply the algebraic Bethe-ansatz method to the transfer matrix
$\Tilde{\tau}(z;\{p\})$ to obtain Bethe eigenstates.
The associated Bethe equations are given by
\begin{align}
\label{eq:Bethe-eq_tilde}
 \prod_{n=1}^{L}\frac{t_{p_{n}}z_{i}^{2}\varepsilon q^{N}-1}
 {t_{p_{n}}z_{i}^{2}q^{2}-\varepsilon q^{N}}
 =\prod_{j(\neq i)}
  \frac{z_{i}^{2}q^{2}-z_{j}^{2}}{z_{i}^{2}-z_{j}^{2}q^{2}}.
\end{align}

The transfer matrix $\tau(z^{2};\{p\})$ of 
the $\tau_{2}$-model defined in \eqref{eq:T-op_tau-2} 
is equivalent to the transfer matrix $\Tilde{\tau}(z;\{p\})$ 
at $\varepsilon q^{N}=1$. 
We set $\omega=q^{2}$
with the primitive $N$th root of unity $q$ for odd $N$ and 
the primitive $2N$th root of unity $q$ for even $N$,
and take $\alpha=x_{p_{i}}^{\frac{1}{2}}y_{p_{i}}^{-\frac{1}{2}}$.
Then, in terms of the operators $Z_{i}$ and $X_{i}$,
the representation of the quantum algebra $U_{q}(\mathfrak{sl}_{2})$
is expressed as
\[
 (k)_{i}=q^{-1}Z_{i}^{-1},
 \quad
 (e)_{i}=
 \frac{x_{p_{i}}^{\frac{1}{2}}y_{p_{i}}^{-\frac{1}{2}}}{q-q^{-1}}
 X_{i}^{-1}(Z_{i}^{-\frac{1}{2}}-Z_{i}^{\frac{1}{2}}),
 \quad
 (f)_{i}=
 \frac{x_{p_{i}}^{-\frac{1}{2}}y_{p_{i}}^{\frac{1}{2}}}{q-q^{-1}}
 (Z_{i}^{\frac{1}{2}}-Z_{i}^{-\frac{1}{2}})X_{i},
\]
by which the $L$-operator $\Tilde{\mathcal{L}}_{i}(z)$
at $\varepsilon q^{N}=1$ takes the form
\begin{align}
 \Tilde{\mathcal{L}}_{i}(z)
 =\begin{pmatrix}
   q^{-\frac{1}{2}}
   \big(-zq^{-\frac{1}{2}}Z^{-\frac{1}{2}}_{i}
   +z^{-1}q^{\frac{1}{2}}Z^{\frac{1}{2}}_{i}\big)
   &x^{-\frac{1}{2}}_{p_{i}}y^{\frac{1}{2}}_{p_{i}}
    (Z^{\frac{1}{2}}_{i}-Z^{-\frac{1}{2}}_{i})X_{i} \\
    x^{\frac{1}{2}}_{p_{i}}y^{-\frac{1}{2}}_{p_{i}}X_{i}^{-1}
    (Z^{-\frac{1}{2}}_{i}-Z^{\frac{1}{2}}_{i})
   &q^{\frac{1}{2}}
   \big(-zq^{\frac{1}{2}}Z^{\frac{1}{2}}_{i}
   +z^{-1}q^{-\frac{1}{2}}Z^{-\frac{1}{2}}_{i}\big) \\
  \end{pmatrix}
 .\nn
\end{align}
The $L$-operator $\Tilde{\mathcal{L}}_{i}(z)$ 
is transformed to the $L$-operator 
$\mathcal{L}_{i}(z^{2};p_{i},\Bar{p}_{i})$ 
defined in \eqref{eq:L-op_SCP} as follows:
\begin{align}
&\begin{pmatrix}
  1 & 0\\
  0 & z^{-1}q^{\frac{1}{2}} \\
 \end{pmatrix}
 t_{p_{i}}^{\frac{1}{2}}zq^{\frac{1}{2}}Z_{i}^{\frac{1}{2}}
 \Tilde{\mathcal{L}}_{i}
 \big(t_{p_{i}}^{\frac{1}{2}}zq^{\frac{1}{2}}\big)
 \begin{pmatrix}
  1 & 0\\
  0 & zq^{-\frac{1}{2}} \\
 \end{pmatrix}
 =\mathcal{L}_{i}(z^{2};p_{i},\Bar{p}_{i}).
 \nn
\end{align}
Through the gauge transformation, 
the Yang-Baxter relation with the $R$-matrix $R_{\mathrm{6v}}(z)$
for the $L$-operator $\Tilde{\mathcal{L}}_{i}(z)$ is 
transformed to the Yang-Baxter relation~\eqref{eq:RTT=TTR}
with the $R$-matrix $R(z)$~\eqref{eq:R-matrix_hom}.
In the case of odd $N$,
the $L$-operator $\Tilde{\mathcal{L}}_{i}(z)$
satisfies the Yang-Baxter relation~\eqref{eq:SLL-relation_CP}.
On the other hand, in the case of even $N$,
the $L$-operator $\Tilde{\mathcal{L}}_{i}(z)$
does not satisfy the relation~\eqref{eq:SLL-relation_CP}
due to the multiplication by the operator $Z_{i}^{\frac{1}{2}}$.
However, the conserved operators derived from an expansion 
of the logarithm of the transfer matrix $\Tilde{\tau}(z;\{p\})$ 
commute with the transfer matrix $t(q_{1},q_{2};\{p\})$
since the operators $Z_{i}^{\frac{1}{2}}$
are canceled out in the derivation.
Furthermore, the product 
$Z_{1}^{\frac{1}{2}}\cdots Z_{L}^{\frac{1}{2}}$,
which appears in each entry of the monodromy matrix 
$\Tilde{\mathcal{T}}(z;\{p\})$,
acts as the constant $q^{M}$ on the sector spanned by
the vectors
$v_{\sigma_{1}}\otimes\cdots\otimes v_{\sigma_{L}}$
satisfying $\sigma_{1}+\cdots+\sigma_{L}=M$.
As we shall see below, each Bethe eigenstate and
its $L(\mathfrak{sl}_{2})$-descendant state belong
to one of the sectors.
Therefore, transfer matrices 
$\Tilde{\tau}(z;\{p\})$ and $\tau(z;\{p\})$
thus share a set of common eigenvectors.

\mathversion{bold}
\subsection{The $\mathfrak{sl}_{2}$ loop algebra symmetry}
\mathversion{normal}
\label{sec:sl_2-loop-algebra}

We now show the $\mathfrak{sl}_{2}$ loop algebra 
$L(\mathfrak{sl}_{2})$ symmetry of the $\tau_{2}$-model. 
We first obtain a representation of the quantum affine algebra 
$U_{q}^{\prime}(\mathfrak{sl}_{2})$ in a limit of 
the entries of the monodromy matrix 
$\Tilde{\mathcal{T}}(z;\{p\})$:
\begin{align}
&A\Define
 \lim_{z\to\infty}\frac{\Tilde{A}(z)}{m(z)q^{-\frac{L}{2}}}
 =\lim_{z\to 0}\frac{\Tilde{D}(z)}{m(z)q^{\frac{L}{2}}}
 =k^{\frac{1}{2}}\otimes\cdots\otimes k^{\frac{1}{2}},
 \nn\\
&B_{\pm}\Define
 \lim_{z^{\pm 1}\to\infty}\frac{\Tilde{B}(z)}
 {m(z)n_{\pm}(z)}
 =\sum_{i=1}^{L}q^{\frac{L+1}{2}-i}
 (t_{p_{i}}^{\mp\frac{1}{2}}q^{\mp\frac{1}{2}})
 \underbrace{
 k^{\pm\frac{1}{2}}\otimes\cdots\otimes 
 k^{\pm\frac{1}{2}}
 }_{i-1}
 \otimes f\otimes
 \underbrace{
 k^{\mp\frac{1}{2}}\otimes\cdots\otimes
 k^{\mp\frac{1}{2}}
 }_{L-i},
 \nn\\
&C_{\pm}\Define
 \lim_{z^{\pm 1}\to\infty}\frac{\Tilde{C}(z)}
 {m(z)n_{\pm}(z)}
 =\sum_{i=1}^{L}q^{-\frac{L+1}{2}+i}
 (t_{p_{i}}^{\mp\frac{1}{2}}q^{\mp\frac{1}{2}})
 \underbrace{
 k^{\mp\frac{1}{2}}\otimes\cdots\otimes 
 k^{\mp\frac{1}{2}}
 }_{i-1}
 \otimes e\otimes
 \underbrace{
 k^{\pm\frac{1}{2}}\otimes\cdots\otimes 
 k^{\pm\frac{1}{2}}
 }_{L-i}, 
 \nn
\end{align}
where $m(z)=\prod_{i=1}^{L}
(t_{p_{i}}^{\frac{1}{2}}zq^{\frac{1}{2}}
-t_{p_{i}}^{-\frac{1}{2}}z^{-1}q^{-\frac{1}{2}})$
and $n_{\pm}(z)=\pm z^{\mp 1}(q-q^{-1})$.
They indeed give a finite-dimensional representation
of $U_{q}^{\prime}(\Hat{\mathfrak{sl}}_{2})$ 
through the map $\pi^{(L)}:
U_{q}^{\prime}(\Hat{\mathfrak{sl}}_{2})
\to(\mathbb{C}^{N})^{\otimes L}$ defined by
\[
 \pi^{(L)}: k_{0,1}, e_{0}, e_{1}, f_{0}, f_{1}
 \mapsto A^{\mp 2}, B_{+}, C_{+}, C_{-}, B_{-},
\]
where $\{k_{i}, e_{i}, f_{i}|i=0,1\}$ is a set of
the Chevalley generators of 
$U_{q}^{\prime}(\Hat{\mathfrak{sl}}_{2})$. 

Second we show that, in the limit $\varepsilon q^{N}\to 1$, 
the representation $\pi^{(L)}$ of the quantum 
affine algebra $U_{q}^{\prime}(\Hat{\mathfrak{sl}}_{2})$
gives a finite-dimensional representation of 
a Borel subalgebra of $L(\mathfrak{sl}_{2})$.
The $\mathfrak{sl}_{2}$ loop algebra is realized 
by the Drinfeld generators 
$\{h_{n},x_{n}^{+},x_{n}^{-}|n=0,1,2,\ldots\}$
satisfying
\begin{align}
 [h_{n},h_{m}]=0,\qquad
 [h_{n},x_{m}^{\pm}]=\pm 2x_{n+m}^{\pm},\qquad
 [x_{n}^{+},x_{m}^{-}]=h_{n+m}. \nn
\end{align}
The algebra has two Borel subalgebras $\mathfrak{b}_{+}$ 
generated by
$\{h_{n},x_{n}^{+},x_{m}^{-}|n\geq 0, m>0\}$
and $\mathfrak{b}_{-}$ generated by
$\{h_{-n},x_{-m}^{+},x_{-n}^{-}|n\geq 0, m>0\}$.
Define the operators
\begin{align}
&H^{(N)}\Define\frac{1}{N}
 \sum_{i=1}^{L}\id\otimes\cdots\otimes\id\otimes
 h\otimes\id\otimes\cdots\otimes\id,
 \nn\\
&B_{\pm}^{(n)}\Define\lim_{\varepsilon q^{N}\to 1}
 \frac{(B_{\pm})^{n}}{[n]!},\qquad
 C_{\pm}^{(n)}\Define\lim_{\varepsilon q^{N}\to 1}
 \frac{(C_{\pm})^{n}}{[n]!}\quad\text{for odd }N,
 \nn
\end{align}
where $hv_{\sigma}=(N-1-2\sigma)v_{\sigma}$ and
$[n]!=[n][n-1]\cdots[1]$.
The operators $B_{\pm}^{(N)}$
and $C_{\pm}^{(N)}$ are well-defined
in the limit $\varepsilon q^{N}\to 1$ since both the operators
$(B_{\pm})^{N}$ and $(C_{\pm})^{N}$ include 
the factor $[N]!$.
They satisfy the relations
\begin{align}
 &[B_{+}^{(N)},B_{-}^{(N)}]=
  [C_{+}^{(N)},C_{-}^{(N)}]=0,\nn\\
 &[H^{(N)},B_{\pm}^{(N)}]=-2B_{\pm}^{(N)},\qquad
  [H^{(N)},C_{\pm}^{(N)}]=2C_{\pm}^{(N)},\nn\\
 &[B_{\pm}^{(N)},[B_{\pm}^{(N)},
  [B_{\pm}^{(N)},C_{\pm}^{(N)}]]]=0,\qquad
  [C_{\pm}^{(N)},[C_{\pm}^{(N)},
  [C_{\pm}^{(N)},B_{\pm}^{(N)}]]]=0. \nn
\end{align}
Here the last two relations are obtained
from the limit $\varepsilon q^{N}\to 1$ of the higher-order
$q$-Serre relations in 
$U_{q}^{\prime}(\Hat{\mathfrak{sl}}_{2})$~\cite{Lusztig_93}.
Then we find that
the map $\varphi_{+}: \mathfrak{b}_{+}\to
\mathrm{End}((\mathbb{C}^{N})^{\otimes L})$ defined by
\[
 \varphi_{+}(h_{0})\Define H^{(N)},\quad
 \varphi_{+}(x_{0}^{+})\Define C_{+}^{(N)},\quad
 \varphi_{+}(x_{1}^{-})\Define B_{+}^{(N)}
\]
is extended to a finite-dimensional representation of
the Borel subalgebra $\mathfrak{b}_{+}$
and the map $\varphi_{-}: \mathfrak{b}_{-}\to
\mathrm{End}((\mathbb{C}^{N})^{\otimes L})$ defined by
\[
 \varphi_{-}(h_{0})\Define H^{(N)},\quad
 \varphi_{-}(x_{-1}^{+})\Define C_{-}^{(N)},\quad
 \varphi_{-}(x_{0}^{-})\Define B_{-}^{(N)}
\]
is also extended to that of
the Borel subalgebra $\mathfrak{b}_{-}$.

\begin{prop}
The $\tau_{2}$-model in a sector specified below
has the Borel subalgebra symmetry in the following sense:
the transfer matrix $\Tilde{\tau}(z)=\Tilde{\tau}(z;\{p\})$ 
at $\varepsilon q^{N}=1$
satisfies
\[
 \big[\Tilde{\tau}(1)^{-1}\Tilde{\tau}(z),
  \varphi_{+}(x)\big]=0 \quad
 \text{for } x\in\mathfrak{b}_{+}
\]
in the sector with $A^{2}=q^{L}$ and
\[
 \big[\Tilde{\tau}(1)^{-1}\Tilde{\tau}(z),
  \varphi_{-}(x)\big]=0 \quad
 \text{for } x\in\mathfrak{b}_{-}
\]
in the sector with $A^{2}=q^{-L}$.
\label{prop:Borel}
\end{prop}
\begin{proof}
In the limit $\varepsilon q^{N}\to 1$, we have
\begin{align}
 &\Tilde{A}(z)B^{(N)}_{\pm}
 =\varepsilon B^{(N)}_{\pm}\Tilde{A}(z)
  -z^{\pm 1}q^{-\frac{L}{2}}B^{(N-1)}_{\pm}
   \Tilde{B}(z)A^{\pm 1},\nn\\
 &\Tilde{D}(z)B^{(N)}_{\pm}
 =\varepsilon B^{(N)}_{\pm}\Tilde{D}(z)
  +z^{\pm 1}q^{\frac{L}{2}}B^{(N-1)}_{\pm}
   \Tilde{B}(z)A^{\mp 1},\nn\\
 &\Tilde{A}(z)C^{(N)}_{\pm}
 =\varepsilon C^{(N)}_{\pm}\Tilde{A}(z)
  +z^{\pm 1}q^{-\frac{L}{2}}C^{(N-1)}_{\pm}
   \Tilde{C}(z)A^{\pm 1},\nn\\
 &\Tilde{D}(z)C^{(N)}_{\pm}
 =\varepsilon C^{(N)}_{\pm}\Tilde{D}(z)
  -z^{\pm 1}q^{\frac{L}{2}}C^{(N-1)}_{\pm}
   \Tilde{C}(z)A^{\mp 1}. \nn
\end{align}
By considering them in the sector with $A^{2}=q^{\pm L}$,
we prove the proposition.
\end{proof}

Let us consider the condition $A^{2}=q^{\pm L}$ in detail.
{}From the relation $A^{2}=k\otimes\cdots\otimes k$,
we have $A^{2}=\varepsilon^{L}q^{(N-1)L-2M}=q^{-L-2M}$
in the sector spanned by the vectors
$v_{\sigma_{1}}\otimes\cdots\otimes v_{\sigma_{L}}$
satisfying $\sigma_{1}+\cdots+\sigma_{L}=M$.
Then the condition $A^{2}=q^{L}$ means $M+L\equiv 0\mod N$
and $A^{2}=q^{-L}$ means $M\equiv 0\mod N$.
One notices that the reference state $|0\rangle$
belongs to the sector with $A^{2}=q^{-L}$.

We now show the $\mathfrak{sl}_{2}$ loop algebra symmetry 
of the $\tau_{2}$-model. 
It is known  that every finite-dimensional irreducible 
representation of the Borel subalgebra $\mathfrak{b}_{\pm}$
is extended to that of the $\mathfrak{sl}_{2}$ loop 
algebra~\cite{Benkart-Terwilliger_04JA,Deguchi_07math-phys}.
Therefore, it follows from Proposition~\ref{prop:Borel} that  
the transfer matrix of the $\tau_{2}$-model has the 
$\mathfrak{sl}_{2}$ loop algebra symmetry. 

Third we now show  that any given regular Bethe state $|R\rangle$ 
in the sector with $A^{2}=q^{\pm L}$ is 
a highest weight vector with respect to 
the representation $\varphi_{\pm}$
of the Borel subalgebra $\mathfrak{b}_{\pm}$
and the highest weight representation generated 
by the Bethe state is irreducible.
A vector $\Omega$ is called highest weight of
the Borel subalgebra $\mathfrak{b}_{+}$
if it is annihilated by $x_{n}^{+}, (n\geq 0)$
and is diagonalized by $h_{n}, (n\geq 0)$,
and is called highest weight of $\mathfrak{b}_{-}$
if it is annihilated by $x_{-n}^{+}, (n>0)$
and is diagonalized by $h_{-n}, (n\geq 0)$.
The conditions are equivalent to~\cite{Deguchi_04JPA}
\begin{align}
&x_{0}^{+}\Omega=0,\qquad
 h_{0}\Omega=r\Omega,
 \qquad
 \frac{(x_{0}^{+})^{m}}{m!}
 \frac{(x_{1}^{-})^{m}}{m!}\Omega=\chi^{+}_{m}\Omega,
 \quad
 (m\in\mathbb{Z}_{>0})
 \qquad
 \text{for } \mathfrak{b}_{+}
 \nn\\
&x_{-1}^{+}\Omega=0,\qquad
 h_{0}\Omega=r\Omega,
 \qquad
 \frac{(x_{-1}^{+})^{m}}{m!}
 \frac{(x_{0}^{-})^{m}}{m!}\Omega=\chi^{-}_{m}\Omega,
 \quad
 (m\in\mathbb{Z}_{>0})
 \qquad
 \text{for } \mathfrak{b}_{-},
 \nn
\end{align}
where $r\in\mathbb{Z}_{>0}$ and 
$\chi^{\pm}_{m}\in\mathbb{C}$.
By using the set $\{\chi^{\pm}_{m}\}$
for a highest weight vector of 
the Borel subalgebra $\mathfrak{b}_{\pm}$,
we define the highest weight polynomial as 
\cite{Deguchi_06DGP,Deguchi_07JSM} 
\[
 P_{\text{D}}^{\pm}(\zeta)
 =\sum_{m\geq 0}\chi^{\pm}_{m}(-\zeta)^{m}.
\]

\begin{prop}
\label{prop:HW-polynomial}
At $\varepsilon q^{N}=1$, 
every regular Bethe state $|R\rangle$
in the sector with $A^{2}=q^{\pm L}$ 
is a highest weight vector with respect to
the representation $\varphi_{\pm}$ of 
the Borel subalgebra $\mathfrak{b}_{\pm}$.
The highest weight polynomial is given by
\begin{align}
\label{eq:HW-poly}
 P^{\pm}_{\mathrm{D}}(\xi^{N})
 =\frac{1}{N}
 \sum_{\tau=0}^{N-1}
 \left(\prod_{n=1}^{L}
 \frac{1-t_{p_{n}}^{\mp N}\xi^{N}}
 {1-t_{p_{n}}^{\mp 1}\xi q^{\pm 2\tau}}\right)
 \frac{1}
 {F_{\pm}(\xi q^{\pm 2\tau})
  F_{\pm}(\xi q^{\pm 2(\tau+1)})},
\end{align}
where $F_{\pm}(\xi)=\prod_{i=1}^{R}(1-\xi z_{i}^{\pm 2})$
and $\{z_{i}\}$ is a regular solution of
the Bethe equations~\eqref{eq:Bethe-eq_tilde}.

\end{prop}

We shall give a proof of Proposition~\ref{prop:HW-polynomial}
in Appendix~\ref{sec:proof_HW-poly}.

Here we can directly show that every highest weight vector of the 
Borel subalgebra becomes a highest weight vector 
of the $\mathfrak{sl}_{2}$ loop algebra in a finite-dimensional 
highest weight representation 
(see, Appendix A of \cite{Deguchi_07JPA}).

Let us discuss a physical consequence of 
generic inhomogeneous parameters. 
For a given regular Bethe state, 
the zeros of polynomial 
$P^{\pm}_{\mathrm{D}}(\zeta)$~\eqref{eq:HW-poly}
should be distinct,  if inhomogeneous parameters $\{p_{n}\}$ 
on the Fermat curve~\eqref{eq:Fermat} are given by generic values. 
If they are distinct, it therefore follows that 
the highest weight representation generated by the regular Bethe state 
is irreducible and the polynomial $P^{\pm}_{\mathrm{D}}(\zeta)$ 
is identified with the Drinfeld polynomial~\cite{Chari-Pressley_91CMP,%
Drinfeld_88SMD,Deguchi_06DGP,Deguchi_07JSM}. 
Assuming that the zeros of the Drinfeld polynomial are distinct, 
we express the distinct zeros $P^{\pm}_{\mathrm{D}}(\zeta)$ by 
$\zeta_{i}, (i=1,2,\ldots,r=\deg P^{\pm}_{\mathrm{D}}(\zeta))$.  
Then, the representation is isomorphic to the tensor product 
of two-dimensional evaluation representations,
$V_{1}(\zeta_{1})\otimes\cdots\otimes V_{1}(\zeta_{r})$,
and the $\tau_{2}$-model in the sector $A^{2}=q^{\pm L}$ 
has the $2^{r}$-dimensional degenerate eigenspace
associated with the regular Bethe state.

\mathversion{bold}
\subsection{Complete $N$-strings and degenerate eigenvectors  
of the $\mathfrak{sl}_{2}$ loop algebra }
\mathversion{normal}
\label{sec:complete-N-strings}

In Propositions \ref{prop:Borel} and \ref{prop:HW-polynomial} 
of Section \ref{sec:sl_2-loop-algebra},
it has been shown in the sector that the $\tau_2$-model has 
the $\mathfrak{sl}_{2}$ loop algebra $L(\mathfrak{sl}_{2})$ symmetry 
and also that every regular Bethe state $|R \rangle$
is a highest weight vector of $L(\mathfrak{sl}_{2})$. 
Therefore, the degenerate eigenspace of the 
$\tau_2$-model associated with the regular Bethe state $|R \rangle$ 
is given by the highest weight representation 
generated by $|R \rangle$ through  
generators of $L(\mathfrak{sl}_{2})$. 

Let us define a complete $N$-string by the set 
$\{\e^{\Lambda}\omega^{-l}|l=1,2, \ldots, N\}$,
where we call $\Lambda$ the center of 
the string~\cite{Fabricius-McCoy_01JSPa}.
By adding $m$ complete $N$-strings 
$\{\e^{\Lambda_{j}}\omega^{-l}|l=1,2,\ldots,N, j=1,2,\ldots,m\}$ 
to a regular solution $\{z_{i}|i=1,2,\ldots,R\}$ 
of the Bethe equations~\eqref{eq:Bethe-eq}
and taking the limit $\Lambda_{j}\to\pm\infty$, 
we obtain a formal solution 
$\{z_{i}\}\cup\{\e^{\Lambda_{j}}\omega^{-l}\}$
of the Bethe equations~\eqref{eq:Bethe-eq} with $M=R+mN$.
We call it a non-regular solution. 
It is clear that the transfer-matrix eigenvalue \eqref{eq:tau-eigenvalue_tau-2}
for a non-regular solution 
$\{z_{i}\}\cup\{\e^{\Lambda_{j}}\omega^{-l}\}$ is the same as
that of the original regular solution $\{z_{i}\}$. 

We now discuss that the SCP transfer matrix $t(q,\Bar{q}(s);\{p\})$
with $q=(x_{q},y_{q},\mu_{q})$ 
and $\Bar{q}(s)=(y_{q},x_{q}\omega^{s},\mu_{q}^{-1})$
should have degenerate eigenspaces. 
We observe that the eigenvalue \eqref{eq:t-eigenvalue-R_SCP} of
the SCP transfer matrix $t(q,\Bar{q}(s);\{p\})$
with a non-regular solution 
$\{z_{i}\}\cup\{\e^{\Lambda_{j}}\omega^{-l}\}$ is the same
as that with the original regular solution $\{z_{i}\}$.
As a consequence, the degenerate eigenspace of 
the transfer matrix $\tau(z;\{p\})$,
which contains a regular Bethe state
and non-regular Bethe states, corresponds to
a degenerate eigenspace of
the transfer matrix $t(q,\Bar{q}(s);\{p\})$.

Non-regular Bethe eigenstates with complete $N$-strings 
may vanish as we shall see in Section~\ref{sec:sl_2-loop-algebra}.
However,  there are several approaches 
to obtain non-zero eigenstates corresponding to 
non-regular solutions such as complete $N$-strings 
\cite{Deguchi-Fabricius-McCoy_01JSP,Fabricius-McCoy_01MPOdyssey,%
Deguchi_02JPA,Deguchi_02IJMPB,Deguchi_07JPA}. 
Thus, from the observation that 
the eigenvalue \eqref{eq:t-eigenvalue-R_SCP} 
does not depend on complete $N$-strings,  
we suggest that the SCP transfer matrix $t(q,\Bar{q}(s);\{p\})$ 
is also degenerate in the $L(\mathfrak{sl}_{2})$-degenerate eigenspace 
of the $\tau_2$-model generated by a regular Bethe state.   

We thus propose a conjecture 
that the $\mathfrak{sl}_{2}$ loop algebra symmetry 
of the $\tau_{2}$-model gives a degenerate eigenspace 
of the SCP transfer matrix $t(q,\Bar{q}(s);\{p\})$ in the sector. 
Intuitively, in terms of complete $N$-strings, we may interpret that 
every $L(\mathfrak{sl}_{2})$-descendant state 
of a given regular Bethe state should be expressed as some linear combination 
of such non-regular Bethe states consisting of complete $N$-strings.   
Furthermore, the Drinfeld polynomial 
$P^{\pm}_{\mathrm{D}}(\zeta)$~\eqref{eq:HW-poly}
is identical to the SCP polynomial 
$P_{\mathrm{SCP}}(\zeta)$~\eqref{eq:SCP-poly} 
associated with the regular Bethe state.  
Thus, the $L(\mathfrak{sl}_{2})$-degenerate 
eigenspace of the $\tau_{2}$-model should have 
exactly the same dimensions as the invariant subspace 
associated with the Ising-like spectrum~\eqref{eq:T-form} 
characterized by the SCP polynomial $P_{\mathrm{SCP}}(\zeta)$. 

\mathversion{bold}
\subsection{The $\mathfrak{sl}_{2}$ loop algebra degeneracy %
and the Ising-like spectrum}
\mathversion{normal}

We now discuss an important consequence of the commutativity of 
the SCP transfer matrix $t(q,\Bar{q}(s);\{p\})$ 
with the transfer matrix $\tau(z,\{p\})$ of the $\tau_{2}$-model.  
Here we note that basis vectors diagonalizing commuting  
transfer matrices do not depend on the spectral parameters.  

We define the completeness of the Bethe ansatz of 
the $\tau_{2}$-model 
at the superintegrable point by the following conjecture: 
\begin{conj}
\label{conj:completeness}
All regular Bethe states 
in the sector with $A^{2}=q^{\pm L}$ 
and their descendants with respect to 
the $\mathfrak{sl}_2$ loop algebra give the complete set of 
the Hilbert space in the sector 
on which  transfer matrix $\tau(z,\{p\})$ 
of the $\tau_2$-model acts. Here we recall $\varepsilon q^{N}=1$. 
\end{conj} 

For generic values of spectral parameter $z$,  
regular Bethe states in the sector are non-degenerate with 
respect to the eigenvalue of transfer matrix $\tau(z,\{p\})$.   
The degeneracy in the eigenspectrum of transfer matrix 
$\tau(z,\{p\})$ should be given only by 
the $\mathfrak{sl}_{2}$ loop algebra symmetry. 
Similarly, for generic spectral parameters, 
regular Bethe states are non-degenerate 
with respect to the eigenvalue 
of the SCP transfer matrix $t(q,\Bar{q}(s);\{p\})$.  
The eigenvalue of $t(q,\Bar{q}(s);\{p\})$ is also generic with respect 
to the spectral parameters, as shown in~\eqref{eq:t-eigenvalue-R_SCP}. 

Thus, if Conjecture \ref{conj:completeness} is valid, i.e. 
the completeness of the Bethe ansatz for the $\tau_2$-model is valid,   
we have the following corollary: 
\begin{cor}
In the sector with $A^{2}=q^{\pm L}$,
the SCP transfer matrix $t(q_1, q_2;\{p\})$ is block-diagonalized 
with respect to the $L(\mathfrak{sl}_{2})$-degenerate 
eigenspaces of the $\tau_{2}$-model associated 
with the regular Bethe states.  Here we recall  $\varepsilon q^{N}=1$. 
\end{cor}

Assuming the arguments for deriving the formula of eigenvalues of  
the diagonal-to-diagonal transfer matrices 
$T_{\text{D}}(x_{q},y_{q})$ and $\Hat{T}_{\text{D}}(x_{q},y_{q})$, 
we have the following conjecture: 
\begin{conj}
\label{conj:main}
In the $L(\mathfrak{sl}_{2})$-degenerate 
eigenspace of the $\tau_{2}$-model associated with a regular 
Bethe state $|R\rangle$, 
the diagonal-to-diagonal transfer matrices 
$T_{\text{D}}(x_{q},y_{q})$ 
and $\Hat{T}_{\text{D}}(x_{q},y_{q})$ of the SCP model
have the Ising-like spectrum~\eqref{eq:T-form} 
associated with the regular Bethe state $|R\rangle$. 
\end{conj}

Let us consider some examples of 
the invariant subspace of the Ising-like spectrum 
associated with a regular Bethe state $|R\rangle$. 
If the degree of the Drinfeld polynomial 
$P^{\pm}_{\mathrm{D}}(\zeta)$ 
is zero, then $|R\rangle$ is an eigenvector of 
both of the two diagonal-to-diagonal transfer matrices 
$T_{\text{D}}(x_{q},y_{q})$ and 
$\hat{T}_{\text{D}}(x_{q},y_{q})$. 
The Bethe state should  generate a singlet 
of the $\mathfrak{sl}_{2}$ loop algebra, 
i.e. a one-dimensional highest weight representation.
One notices that, if $R=L(N-1)/2$,  
the degree of the Drinfeld polynomial 
$P^{\pm}_{\mathrm{D}}(\zeta)$ is zero.  

However, if the degree of the Drinfeld polynomial 
is nonzero and given by $r$, 
the SCP transfer matrix $t(q_{1},q_{2};\{p\})$ 
should be block-diagonalized 
at least with respect to the $L(\mathfrak{sl}_{2})$-degenerate 
eigenspace of the $\tau_2$-model associated with $|R\rangle$. 
Furthermore, the SCP transfer matrix $t(q,\Bar{q}(s);\{p\})$
 should be degenerate in the $2^{r}$-dimensional 
$L(\mathfrak{sl}_{2})$-degenerate eigenspace 
of the $\tau_2$-model associated with $|R\rangle$, 
as it was conjectured in Section 4.3.

\mathversion{bold}
\subsection{$N=2$ case}
\mathversion{normal}

We verify in the case of $N=2$ with a set of 
homogeneous parameters 
that the Hamiltonian of the SCP model
has the Ising-like spectrum
in the $L(\mathfrak{sl}_{2})$-degenerate 
eigenspace of the $\tau_{2}$-model. 
In the case, the SCP model is the two-dimensional Ising model.
The Hamiltonians of the SCP model
and the $\tau_{2}$-model are given in the forms
\[
 H_{\text{SCP}}=\sum_{i=1}^{L}\sigma_{i}^{z}
 +\lambda\sum_{i=1}^{L}\sigma_{i}^{x}\sigma_{i+1}^{x},\qquad
 H_{\tau_{2}}=\sum_{i=1}^{L}
 (\sigma_{i}^{x}\sigma_{i+1}^{y}-\sigma_{i}^{y}\sigma_{i+1}^{x}),
\]
where $\sigma^{x}$, $\sigma^{y}$ and $\sigma^{z}$
are Pauli's matrices.
In terms of Jordan-Wigner's fermion operators:
\[
 c_{i}=\sigma_{i}^{+}\prod_{j=1}^{i-1}\sigma_{j}^{z},\qquad
 \Tilde{c}_{k}=\frac{1}{L}\sum_{i=1}^{L}
 \e^{-\sqrt{-1}(ki+\frac{\pi}{4})}c_{i},
\]
the Hamiltonian $H_{\tau_{2}}$ 
in the sector with $S^{z}\Define\frac{1}{2}\sum_{i=1}^{L}
\sigma_{i}^{z}\equiv 0\mod 2$ is written as
\[
 H_{\tau_{2}}=\sum_{k\in K}\sin(k)
 \Tilde{c}^{\dagger}_{k}\Tilde{c}_{k},
\]
where $K=\{\frac{\pi}{L},\frac{3\pi}{L},
\ldots,\frac{(L-1)\pi}{L}\}$.
For even $L$, the $\mathfrak{sl}_{2}$ loop algebra symmetry 
describing a degenerate eigenspace of the $\tau_{2}$-model
is given by 
\[
 h_{n}=\sum_{k\in K}\cot^{2n}\Big(\frac{k}{2}\Big)(H)_{k},
 \quad
 x_{n}^{+}=\sum_{k\in K}\cot^{2n+1}\Big(\frac{k}{2}\Big)(E)_{k},
 \quad
 x_{n}^{-}=\sum_{k\in K}\cot^{2n-1}\Big(\frac{k}{2}\Big)(F)_{k},
\]
where $\{(H)_{k}, (E)_{k}, (F)_{k}\}$
is a two-dimensional representation of 
the $\mathfrak{sl}_{2}$ algebra given by
\[
 (H)_{k}=1-\Tilde{c}^{\dagger}_{k}\Tilde{c}_{k}
          -\Tilde{c}^{\dagger}_{-k}\Tilde{c}_{-k},\quad
 (E)_{k}=\Tilde{c}_{-k}\Tilde{c}_{k},\quad
 (F)_{k}=\Tilde{c}_{k}^{\dagger}\Tilde{c}_{-k}^{\dagger}.
\]
Here we should remark that 
for the XX model under the periodic boundary conditions 
the Chevalley generators of the $\mathfrak{sl}_{2}$ 
loop algebra symmetry were constructed in terms of the free fermion operators 
\cite{Deguchi-Fabricius-McCoy_01JSP}.

The reference state $|0\rangle$, which is a highest weight vector,
i.e. $x_{n}^{+}|0\rangle=0$ and $h_{n}|0\rangle
=\sum_{k}\cot^{2n}(k/2)|0\rangle$,
generates a $2^{L/2}$-dimensional irreducible representation
corresponding to a degenerate eigenspace of 
the Hamiltonian $H_{\tau_{2}}$.
On the other hand, in the sector,
the Hamiltonian $H_{\mathrm{SCP}}$ is expressed as
\[
 H_{\mathrm{SCP}}=2\sum_{k\in K}(H)_{k}
 -2\lambda \sum_{k\in K}
 \big(\cos(k)(H)_{k}+\sin(k)\big((E)_{k}+(F)_{k}\big)\big).
\]
It is clear that the Hamiltonian $H_{\mathrm{SCP}}$ acts on 
the $2^{L/2}$-dimensional irreducible representation space.
The $2^{L/2}$ eigenvalues of $H_{\mathrm{SCP}}$ 
and the corresponding eigenstates are given by
\begin{align}
&E(K_{+};K_{-})
 =2\sum_{k\in K_{+}}\sqrt{1-2\lambda\cos(k)+\lambda^{2}}
 -2\sum_{k\in K_{-}}\sqrt{1-2\lambda\cos(k)+\lambda^{2}},
 \nn\\
&|K_{+};K_{-}\rangle
 =\prod_{k\in K_{+}}(\cos\theta_{k}+\sin\theta_{k}(F)_{k})
  \prod_{k\in K_{-}}(\sin\theta_{k}-\cos\theta_{k}(F)_{k})
  |0\rangle,
 \nn
\end{align}
where $K_{+}$ and $K_{-}$ are such disjoint subsets
of $K$ that $K=K_{+}\cup K_{-}$
and $\tan(2\theta_{k})
=\frac{\lambda\sin(k)}{\lambda\cos(k)-1}$.

\section*{Acknowledgments}

The authors would like to thank Prof.~A.~Kuniba 
and Prof.~N.~Hatano for helpful comments. 
One of the authors (A.N.) acknowledges support
from Grant-in-Aid for Young Scientists (B) No.~20740217,
and also from Core Research for Evolutional Science and Technology 
of Japan Science and Technology Agency.
The present study is partially supported by 
Grant-in-Aid for Scientific Research (C) No.~17540351.  

\appendix
\mathversion{bold}
\section{Relations among the operators in \eqref{eq:T-op_tau-2}}
\mathversion{normal}

\begin{lm}
\label{lm:ABCD}
Let $A_{i}, B_{i}, C_{i}$ and $D_{i}$
denote $A(z_{i}), B(z_{i}), C(z_{i})$ and $D(z_{i})$, 
respectively. We have
\begin{align}
\label{eq:ABCD-relation}
&A_{0}B_{i_{1}}\cdots B_{i_{n}}
 =\omega^{n}\Bigg(\!\Big(\prod_{p=1}^{n}f_{0i_{p}}\Big)
  B_{i_{1}}\cdots B_{i_{n}}A_{0}
  -\!\sum_{p=1}^{n}
   \Big(\!\prod_{q(\neq p)}\! f_{i_{p}i_{q}}\Big)
   g_{0i_{p}}B_{0}B_{i_{1}}\cdots
   \!\overset{i_{p}}{\Check{\,}}\!
   \cdots B_{i_{n}}A_{i_{p}}
  \Bigg),
 \nn\\
&D_{0}B_{i_{1}}\cdots B_{i_{n}}
 =\omega^{n}\Bigg(\!\Big(\prod_{p=1}^{n}f_{i_{p}0}\Big)
   B_{i_{1}}\cdots B_{i_{n}}D_{0}
  +\!\sum_{p=1}^{n}
   \Big(\!\prod_{q(\neq p)}\! f_{i_{q}i_{p}}\Big)
   g_{0i_{p}}B_{0}B_{i_{1}}\cdots
   \!\overset{i_{p}}{\Check{\,}}\!
   \cdots B_{i_{n}}D_{i_{p}}
  \Bigg),
 \nn\\
&C_{0}A_{i_{1}}\cdots A_{i_{n}}
 =\omega^{n}\Bigg(\!\Big(\prod_{p=1}^{n}f_{i_{p}0}\Big)
   A_{i_{1}}\cdots A_{i_{n}}C_{0}
  +\!\sum_{p=1}^{n}
   \Big(\!\prod_{q(\neq p)}\! f_{i_{q}i_{p}}\Big)
   g_{0i_{p}}A_{0}A_{i_{1}}\cdots
   \!\overset{i_{p}}{\Check{\,}}\!
   \cdots A_{i_{n}}C_{i_{p}}
  \Bigg),
 \nn\\
&C_{0}D_{i_{1}}\cdots D_{i_{n}}
 =\omega^{n}\Bigg(\!\Big(\prod_{p=1}^{n}f_{0i_{p}}\Big)
   D_{i_{1}}\cdots D_{i_{n}}C_{0}
  -\!\sum_{p=1}^{n}
   \Big(\!\prod_{q(\neq p)}\! f_{i_{p}i_{q}}\Big)
   g_{0i_{p}}D_{0}D_{i_{1}}\cdots
   \!\overset{i_{p}}{\Check{\,}}\!
   \cdots D_{i_{n}}C_{i_{p}}
  \Bigg),
 \nn\\
&D_{0}A_{i_{1}}\cdots A_{i_{n}}
 \nn\\
&=A_{i_{1}}\cdots A_{i_{n}}D_{0}
 +\omega^{n}\sum_{p=1}^{n}
  \Big(\!\prod_{q(\neq p)}f_{i_{q}i_{p}}\Big)
  \Big(g_{0i_{p}}B_{0} A_{i_{1}}\cdots
   \!\overset{i_{p}}{\Check{\,}}\!
   \cdots A_{i_{n}}C_{i_{p}}+
  g_{i_{p}0}B_{i_{p}}A_{i_{1}}\cdots
   \!\overset{i_{p}}{\Check{\,}}\!
   \cdots A_{i_{n}}C_{0}
  \Big),
 \nn\\
&C_{0}B_{i_{1}}\cdots B_{i_{n}}
 =\omega^{n}B_{i_{1}}\cdots B_{i_{n}}C_{0}
  \nn\\
&\quad
 +\omega^{2n-1}\Bigg(
  \sum_{p=1}^{n}
  g_{0i_{p}}B_{i_{1}}\cdots
  \!\overset{i_{p}}{\Check{\,}}\!\cdots B_{i_{n}}
  \Big(
  \Big(\prod_{q(\neq p)}f_{0i_{q}}f_{i_{q}i_{p}}\Big)
  A_{0}D_{i_{p}}-
  \Big(\prod_{q(\neq p)}f_{i_{p}i_{q}}f_{i_{q}0}\Big)
  A_{i_{p}}D_{0}
  \Big)
  \nn\\
&\hspace{59pt}
 -\sum_{p\neq q}
  g_{0i_{p}}g_{0i_{q}}B_{0}B_{i_{1}}\cdots
  \!\overset{i_{p}}{\Check{\,}}\!\cdots
  \!\overset{i_{q}}{\Check{\,}}\!\cdots B_{i_{n}}
  f_{i_{p}i_{q}}
  \Big(\!\!\prod_{r(\neq p,q)}\!\!
  f_{i_{p}i_{r}}f_{i_{r}i_{q}}\Big)
  A_{i_{p}}D_{i_{q}}
  \Bigg),
\end{align}
where $f_{ij}=f(z_{i}/z_{j})$
and $g_{ij}=g(z_{i}/z_{j})$ with 
\[
 f(z)=\frac{z-\omega}{(z-1)\omega},\quad
 g(z)=\frac{1-\omega}{(z-1)\omega}.
\]
\end{lm}
\begin{proof}
The relations with $n=1$ are equivalent to
the Yang-Baxter relation~\eqref{eq:RTT=TTR}.
For $n\geq 2$, we employ induction on $n$
with the identity in Lemma~\ref{lm:key-identity_0}.
\end{proof}

\mathversion{bold}
\section{Identities}
\mathversion{normal}

We collect here several useful identities.

\begin{lm}
Let $S$ be a subset of $\Sigma_{R}=\{1,2,\ldots,R\}$.
We then have
\begin{align}
&\prod_{i\in S}
 \Bigg(a_{i}\prod_{j\in\Sigma_{R}\setminus S}f_{ij}\Bigg)
=\prod_{i\in S}
 \Bigg(d_{i}\prod_{j\in\Sigma_{R}\setminus S}f_{ji}\Bigg).
 \nn
\end{align}
\end{lm}

The following three identities of rational functions
are proved by verifying that
all the residues in the left-hand side are zero.

\begin{lm}
\label{lm:key-identity_0}
\[
 \Bigg(\Big(\prod_{i=1}^{n}f_{ik}\Big)
 -\Big(\prod_{i=1}^{n}f_{il}\Big)\Bigg)g_{kl}
 +\sum_{i=1}^{n}\Big(\prod_{j(\neq i)}f_{ij}\Big)
  g_{ki}g_{il}=0.
\]
\end{lm}

Let $\{i_{\ell}\}$, $\{j_{\ell}\}$ and $\{k_{\ell}\}$ 
be such disjoint subsets of 
the set $\Sigma_{n}=\{1,2,\ldots,n\}$ that
$\sharp\{i_{\ell}\}=n_{B}$, 
$\sharp\{j_{\ell}\}=n_{A}$, 
$\sharp\{k_{\ell}\}=n_{D}$ and $n_{B}+n_{A}+n_{D}=n$.
We have the following identities:

\begin{lm}
\label{lm:key-identity_1}
\begin{align}
&\omega^{n_{A}}
 \prod_{p=1}^{n_{A}}f_{j_{p},n}
 -\frac{\mu_{\tau^{\prime}-n_{D}+1,\tau}(z_{n})}
 {\mu_{\tau^{\prime}-n_{D}+1,\tau+n_{A}}(z_{n})}
 \nn\\
&+\sum_{p=1}^{n_{A}}
 \frac{\mu_{\tau^{\prime},\tau+n_{A}-1}(z_{j_{p}})}
 {\mu_{\tau^{\prime}\tau}(z_{j_{p}})}
 \frac{\mu_{\tau^{\prime}-n_{D}+1,\tau}(z_{n})}
 {\mu_{\tau^{\prime}-n_{D}+1,\tau+n_{A}-1}(z_{n})}
 \frac{\nu_{\tau+n_{A}}(z_{n})}{\nu_{\tau+n_{A}}(z_{j_{p}})}
 \omega^{n_{A}}g_{n,j_{p}}\prod_{r=1\atop r(\neq p)}^{n_{A}}
 f_{j_{r},j_{p}}=0.
 \nn
\end{align}
or, explicitly,
\begin{align}
 \prod_{p=1}^{n_{A}}
 \frac{z_{n_{B}+p}\!-\!z_{n}\omega}{z_{n_{B}+p}\!-\!z_{n}}
-\frac{tz_{n}\omega^{\tau+n_{A}+1}\!-\!1}
 {tz_{n}\omega^{\tau+1}\!-\!1}
+\sum_{p=1}^{n_{A}}
 \frac{tz_{n_{B}+p}\omega^{\tau+1}\!-\!1}
 {tz_{n}\omega^{\tau+1}\!-\!1}
 \frac{z_{n}(1\!-\!\omega)}{z_{n}\!-\!z_{n_{B}+p}}
 \prod_{r=1 \atop r(\neq p)}^{n_{A}}
 \frac{z_{n_{B}+r}\!-\!z_{n_{B}+p}\omega}
 {z_{n_{B}+r}\!-\!z_{n_{B}+p}}
 =0.
 \nn
\end{align}
\end{lm}

\begin{lm}
\label{lm:key-identity_2}
\begin{align}
&\sum_{{\{j_{\ell}\}, \{k_{\ell}\} \atop n_{B}+n_{A}+n_{D}=n}}
  \hspace{-5pt}
  \prod_{p=1}^{n_{A}}
  \frac{\mu_{\tau+n_{D},\tau+n_{D}}(z_{j_{p}})}
  {\mu_{\tau+n_{D},\tau+n_{D}+p-1}(z_{j_{p}})}
  \nu_{\tau+n_{D}+p}(z_{j_{p}})
  \prod_{q=1}^{n_{D}}
  \frac{-\mu_{\tau+n_{D},\tau+n_{D}}(z_{k_{q}})}
  {\mu_{\tau+n_{D}-q+1,\tau+n_{D}}(z_{k_{q}})}
  \nu_{\tau+n_{D}-q+1}(z_{k_{q}})
  \nn\\
&\quad\times
  \prod_{i\in\{i_{\ell}\} \atop j\in\{j_{\ell}\}}
  \!\!\omega f_{j_{p}i_{r}}\!\!
  \prod_{i\in\{i_{\ell}\} \atop k\in\{k_{\ell}\}}
  \!\!\omega f_{k_{q}i_{r}}\!\!
  \prod_{j\in\{j_{\ell}\} \atop k\in\{k_{\ell}\}}
  \!\!\omega f_{k_{q}j_{p}}
  =0.
 \nn
\end{align}
or, explicitly,
\begin{align}
&\sum_{{\{j_{\ell}\}, \{k_{\ell}\} \atop n_{A}+n_{D}=n-n_{B}}}
  \hspace{-12pt}(-)^{n_{D}}
  \prod_{p=1}^{n_{A}}
  \frac{1}{tz_{j_{p}}\omega^{\tau+n_{D}+1}-1}
  \prod_{q=1}^{n_{D}}
  \frac{\omega^{q-1}}{tz_{k_{q}}\omega^{\tau+n_{D}}-1}
  \prod_{p=1}^{n_{A}}
  \prod_{q=1}^{n_{D}}
  \frac{z_{k_{q}}-z_{j_{p}}\omega}{z_{k_{q}}-z_{j_{p}}}
  =0.
 \nn
\end{align}
\end{lm}

\section{Proof of Proposition \ref{prop:HW-polynomial}}
\label{sec:proof_HW-poly}
\mathversion{bold}
\mathversion{normal}

We give a proof of Proposition \ref{prop:HW-polynomial}.
The detailed proof for the case of the XXZ-Heisenberg spin chain
at roots of unity is presented in~\cite{Deguchi_07JPA}.
Here we show only some different points from it.

 For simplicity, we consider the representation $\varphi_{+}$
of the Borel subalgebra $\mathfrak{b}_{+}$.
Let $\Tilde{A}_{i}=\Tilde{A}(z_{i})$, 
$\Tilde{B}_{i}=\Tilde{B}(z_{i})$,
$\Tilde{C}_{i}=\Tilde{C}(z_{i})$ and 
$\Tilde{D}_{i}=\Tilde{D}(z_{i})$
for $i\in\Sigma_{M}=\{1,2,\ldots,M\}$. 
One of the relations in Lemma~\ref{lm:ABCD}
is rewritten as follows:
\begin{align}
\label{eq:CBBB_1}
 &\Tilde{C}_{0}\Tilde{B}_{1}\cdots \Tilde{B}_{M} \nn\\
 &=\Tilde{B}_{1}\cdots \Tilde{B}_{M}\Tilde{C}_{0}
  +\sum_{i=1}^{M}\Tilde{g}_{0i}
   \Tilde{B}_{1}\cdots\overset{i}{\Check{\;}}
   \cdots \Tilde{B}_{M}
   \Big(
   \Big(\prod_{j(\neq i)}
   \Tilde{f}_{0j}\Tilde{f}_{ji}\Big)
   \Tilde{A}_{0}\Tilde{D}_{i}
   -\Big(\prod_{j(\neq i)}
   \Tilde{f}_{j0}\Tilde{f}_{ij}\Big)
   \Tilde{A}_{i}\Tilde{D}_{0}
   \Big) \nn\\
 &\quad
  -\sum_{i\neq j}\Tilde{g}_{0i}\Tilde{g}_{0j}
   \Tilde{B}_{0}\Tilde{B}_{1}
   \cdots\overset{i}{\Check{\;}}
   \cdots\overset{j}{\Check{\;}}
   \cdots\Tilde{B}_{M}\Tilde{f}_{ij}
   \Big(
   \prod_{l(\neq i,j)}\Tilde{f}_{il}
   \Tilde{f}_{lj}
   \Big)
   \Tilde{A}_{i}\Tilde{D}_{j},
\end{align}
where
\[
 \Tilde{f}_{ij}=\Tilde{f}(z_{i}/z_{j})
 =\frac{z_{i}^{2}q^{-1}-z_{j}^{2}q}{z_{i}^{2}-z_{j}^{2}},
 \qquad
 \Tilde{g}_{ij}=\Tilde{g}(z_{i}/z_{j})
 =\frac{z_{i}z_{j}(q^{-1}-q)}{z_{i}^{2}-z_{j}^{2}}.
\]

\begin{lm}
Let $S_{n}=\{i_{1},i_{2},\ldots,i_{n}\}$ 
be a subset of the set $\Sigma_{M}$.
We have
\begin{align}
\label{eq:CBBB_2}
 (C_{+})^{n}
 \Big(\prod_{l\in\Sigma_{M}}\Tilde{B}_{l}\Big)|0\rangle
&=\Delta(S_{n};\Sigma_{M})
  \sum_{S_{n}\subset\Sigma_{M}}
  \Big(\prod_{l\in\Sigma_{M}\setminus S_{n}}
  \Tilde{B}_{l}\Big)|0\rangle
\end{align}
with the coefficient $\Delta(S_{n};\Sigma_{M})$ given by
\[
 \Delta(S_{n};\Sigma_{M})
 =\Big(\prod_{i\in S_{n}}z_{i}\Big)
  \sum_{P\in\mathfrak{S}_{n}}
  \sum_{l=0}^{n}(-)^{l}
  \left[{n \atop l}\right]q^{\frac{n(n-1)}{2}-(n-1)l}
  \hspace{-10pt}\prod_{1\leq j\leq n-l}\hspace{-10pt}
  \alpha_{i_{Pj}}^{\Sigma_{M}\setminus S_{n}}
  \hspace{-10pt}\prod_{n-l<j\leq n}\hspace{-10pt}
  \Bar{\alpha}_{i_{Pj}}^{\Sigma_{M}\setminus S_{n}}
  \hspace{-10pt}\sum_{1\leq r<s\leq n}\hspace{-10pt}
  \Tilde{f}_{i_{Pr},i_{Ps}},
\]
Here $\mathfrak{S}_{n}$ is the symmetric group of order $n$
acting on the set $\{1,2,\ldots,n\}$ and
\begin{align}
&\alpha^{S}_{i}=\alpha^{S}(z_{i})
 \Define
  q^{-\frac{N-1}{2}L}
  \prod_{n=1}^{L}
  (t_{p_{n}}^{\frac{1}{2}}z_{i}
   \epsilon^{\frac{1}{2}}q^{\frac{N}{2}}
  -t_{p_{n}}^{-\frac{1}{2}}z_{i}^{-1}
   \epsilon^{-\frac{1}{2}}q^{-\frac{N}{2}})
  \prod_{j\in S}q\Tilde{f}_{ij}, 
 \nn\\
&\Bar{\alpha}^{S}_{i}=\Bar{\alpha}^{S}(z_{i})
 \Define
  q^{\frac{N-1}{2}L}
  \prod_{n=1}^{L}
  (t_{p_{n}}^{\frac{1}{2}}z_{i}
   \epsilon^{-\frac{1}{2}}q^{-\frac{N}{2}+1}
  -t_{p_{n}}^{-\frac{1}{2}}z_{i}^{-1}
   \epsilon^{\frac{1}{2}}q^{\frac{N}{2}-1})
  \prod_{j\in S}q^{-1}\Tilde{f}_{ji}. 
 \nn
\end{align}
\end{lm}
\begin{proof}
The case $n=1$ is obtained in the limit $z_{0}\to\infty$
of the relation~\eqref{eq:CBBB_1} divided 
by the factor $m(z_{0})n_{+}(z_{0})$.
For general $n$, we use induction on $n$.
\end{proof}

We consider a diagonal condition
$x_{0}^{+}x_{1}^{-}\Omega=\chi_{1}^{+}\Omega$
for the regular Bethe state 
$|R\rangle$ in the sector with $A^{2}=q^{L}$.
Set $M=R+N$ and $n=N$ in \eqref{eq:CBBB_2}.
Let $\{z_{i}|1\leq i\leq R\}$ be
a regular solution of 
the Bethe equations~\eqref{eq:Bethe-eq_tilde}
and put $\{z_{R+l}^{2}=\epsilon_{l}^{-1}
\Define \e^{\Lambda}q^{-2l}|1\leq l\leq N\}$.
We assume that the solution $\{z_{i}|1\leq i\leq R\}$
is also regular in the limit $\varepsilon q^{N}\to 1$.
Then we have
\begin{align}
 (C_{+})^{N}
 \Big(\prod_{l\in Z_{N}}\Tilde{B}_{l}\Big)
 \Big(\prod_{i\in\Sigma_{R}}\Tilde{B}_{i}\Big)|0\rangle
&=\Delta(S_{N};\Sigma_{R+N})
  \sum_{S_{n}\subset\Sigma_{R+N}}
  \Big(\prod_{l\in\Sigma_{R+N}\setminus S_{N}}
  \Tilde{B}_{l}\Big)|0\rangle,
 \nn
\end{align}
where $Z_{N}\Define\{R+1,\ldots,R+N\}$.
We investigate the diagonal term $S_{N}=Z_{N}$
in the limit $\Lambda\to -\infty$
after dividing both sides of the equation by the factor 
$\prod_{l\in Z_{N}}m(z_{R+l})n_{+}(z_{R+l})$.

\begin{lm}
We have
\begin{align}
&\Hat{\Delta}(Z_{N};\Sigma_{R+N})\Define
 \Big(\prod_{l\in Z_{N}}
 \frac{1}{m(z_{l})n_{+}(z_{l})}\Big)
 \Delta(Z_{N};\Sigma_{R+N}) 
 \nn\\
&=\epsilon_{0}^{-N}
  \frac{[N]!}{q^{\frac{N(N+1)}{2}+N}(q-q^{-1})^{N}}
  \Phi(\epsilon_{0})
  \sum_{l=0}^{N}(-)^{l}\left[{N \atop l}\right]
  q^{-(N-1)l}
  \frac{\prod_{j=1}^{N-1}
  \phi_{+}(\epsilon_{j+l}\varepsilon q^{-N})}
  {F_{+}(\epsilon_{l})F_{+}(\epsilon_{l+1})}, \nn
\end{align}
where 
\[
 \phi_{+}(\xi)=\prod_{n=1}^{L}(1-t_{n}^{-1}\xi),\quad
 F_{+}(\xi)=\prod_{i=1}^{R}(1-z_{i}^{2}\xi),\quad
 \Phi(\epsilon_{0})\Define
 \frac{\phi_{+}(\epsilon_{0}\varepsilon q^{N})
  F_{+}(\epsilon_{0})F_{+}(\epsilon_{N+1})}
  {\prod_{j=1}^{N}\phi_{+}(\epsilon_{j}q^{-1})}.
\]
\end{lm}
\begin{proof}
{}From $S_{N}=\{i_{1},\ldots,i_{N}\}=\{R+1,\ldots,R+N\}$, 
we have
\begin{align}
&\prod_{l\in Z_{N}}
 \frac{1}{m(z_{l})}
 \sum_{P\in\mathfrak{S}_{N}}
 \prod_{1\leq j\leq N-l}\hspace{-2mm}
 \alpha_{R+Pj}^{\Sigma_{R}}
 \hspace{-2mm}\prod_{N-l<j\leq N}\hspace{-2mm}
 \Bar{\alpha}_{R+Pj}^{\Sigma_{R}}
 \hspace{-2mm}\prod_{1\leq r<s\leq N}\hspace{-2mm}
 \Tilde{f}_{R+Pr,R+Ps}
 \nn\\
&=\hspace{-3pt}\prod_{l<j\leq N}\hspace{-5pt}
 \Big(
 \frac{\phi_{+}(\epsilon_{j}\varepsilon q^{-N})}
 {\phi_{+}(\epsilon_{j}q^{-1})}
 \frac{F_{+}(\epsilon_{j+1})}{F_{+}(\epsilon_{j})}
 \Big)\hspace{-4pt}
 \prod_{1\leq j\leq l}
 \Big(
 \frac{\phi_{+}(\epsilon_{j}\varepsilon q^{N-2})}
 {\phi_{+}(\epsilon_{j}q^{-1})}
 \frac{F_{+}(\epsilon_{j-1})}{F_{+}(\epsilon_{j})}
 \Big)[N]!
 \nn\\
&=\frac{\phi_{+}(\epsilon_{0}\varepsilon q^{N})
  F_{+}(\epsilon_{0})F_{+}(\epsilon_{N+1})}
  {\prod_{j=1}^{N}\phi_{+}(\epsilon_{j}q^{-1})}
  \frac{\prod_{j=1}^{N-1}
  \phi_{+}(\epsilon_{j+l}\varepsilon q^{-N})}
  {F_{+}(\epsilon_{l})F_{+}(\epsilon_{l+1})}[N]!. \nn
\end{align}
Here we have used the fact that
$\prod_{1\leq r<s\leq N}\Tilde{f}_{R+Pr,R+Ps}=0$
unless $P$ is the longest element in the symmetric
group $\mathfrak{S}_{N}$.
\end{proof}

We define $\Tilde{\chi}^{+}_{m}$ 
by the following series expansion:
\begin{align}
\label{eq:def-chi}
 \frac{\prod_{j=1}^{N-1}
 \phi_{+}(\xi\varepsilon q^{2j-N-1})}
 {F_{+}(\xi q)F_{+}(\xi q^{-1})}
 =\sum_{m\geq 0}\Tilde{\chi}^{+}_{m}(-\xi)^{m}.
\end{align}

\begin{lm}
In the limit
$\Lambda\to\infty$, that is, $\epsilon_{0}\to 0$, we have
\begin{align}
 \Hat{\Delta}(Z_{N};\Sigma_{R+N}) 
&=\Tilde{\chi}_{N}^{+}([N]!)^{2}+O(\epsilon_{0}). \nn
\end{align}
\end{lm}
\begin{proof}
Put $\xi=\epsilon_{0}q^{2l+1}$
in the definition of $\chi^{+}_{N}$ \eqref{eq:def-chi}.
Then
\begin{align}
&\sum_{l=0}^{N}(-)^{l}
 \left[{N \atop l}\right]q^{-(N-1)l}
 \frac{\prod_{j=1}^{N-1}
       \phi_{+}(\epsilon_{0}\varepsilon q^{2j+2l-N})}
 {F_{+}(\epsilon_{0}q^{2l})
  F_{+}(\epsilon_{0}q^{2l+2})}
 =\sum_{l=0}^{N}(-)^{l}
  \left[{N \atop l}\right]
  \sum_{m=0}^{\infty}
  \Tilde{\chi}^{+}_{m}\,(-\epsilon_{0}q)^{m}
  q^{(2m-N+1)l}
  \nn\\
&=\sum_{m=0}^{\infty}\Tilde{\chi}^{+}_{m}
  (-\epsilon_{0}q)^{m}
  \prod_{l=0}^{N-1}(1-q^{2(m-l)})
 =\Tilde{\chi}^{+}_{N}\epsilon_{0}^{N}[N]!\;
  q^{\frac{N(N+1)}{2}+N}(q-q^{-1})^{N}
  +O(\epsilon_{0}^{N+1}),
  \nn
\end{align}
where we have used the $q$-binomial theorem
and $\prod_{l=0}^{N-1}(1-q^{2(m-l)})=0$
for $0\leq m\leq N-1$.
\end{proof}

\begin{prop}
Let $q$ be the $N$th primitive root of unity for odd $N$ 
and the $2N$th primitive root of unity for odd $N$.
The regular Bethe state $|R\rangle$ 
in the sector with $A^{2}=q^{L}$ satisfies
\[
 \frac{\varphi_{+}(x_{0}^{+})^{m}}{m!}
 \frac{\varphi_{+}(x_{1}^{-})^{m}}{m!}|R\rangle
 =\chi_{mN}^{+}|R\rangle,
\]
where $\chi_{m}^{+}
=\lim_{\varepsilon q^{N}\to 1}\Tilde{\chi}_{m}^{+}$.
\end{prop}
\begin{proof}
We consider only the case $m=1$.
The case of general $m$ is proved in a similar way
by setting $M=R+mN$ and $n=mN$ in \eqref{eq:CBBB_2}.
{}From the lemma above, we have
\[
 (C_{+})^{N}
 \Big(\prod_{l\in Z_{N}}
 \frac{1}{m_{+}(z_{l})n(z_{l})}\Tilde{B}_{l}\Big)
 |R\rangle
 =\chi_{N}^{+}([N]!)^{2}|R\rangle
 +O(\epsilon_{0})
 +\text{off-diagonal terms},
\]
which, in the limit $\Lambda\to\infty$, yields
\[
 \frac{(C_{+})^{N}}{[N]!}\frac{(B_{+})^{N}}{[N]!}
 |R\rangle
 =\Tilde{\chi}_{N}^{+}|R\rangle+\text{off-diagonal terms}.
\]
In the sector with $A^{2}=q^{L}$,
the off-diagonal terms vanish 
in the limit $\varepsilon q^{N}\to 1$~\cite{Deguchi_04JPA}.
\end{proof}

By taking the limit $\varepsilon q^{N}\to 1$ in the definition
of $\Tilde{\chi}_{m}^{+}$~\eqref{eq:def-chi},  we have
\[
 \frac{\prod_{j=1}^{N-1}\phi_{+}(\xi q^{2j-1})}
 {F_{+}(\xi q)F_{+}(\xi q^{-1})}
 =\sum_{m\geq 0}\chi^{+}_{m}(-\xi)^{m}.
\]
The numerator of the left-hand side is rewritten as
\[
 \prod_{j=1}^{N-1}\phi_{+}(\xi q^{2j-1})
 =\prod_{n=1}^{L}\frac{1-t_{p_{n}}^{-N}\xi^{N}q^{-N}}
 {1-t_{p_{n}}^{-1}\xi q^{-1}}.
\]
Then we obtain
\[
 \Big(\prod_{n=1}^{L}\frac{1-t_{p_{n}}^{-N}\xi^{N}q^{-N}}
 {1-t_{p_{n}}^{-1}\xi q^{-1}}\Big)
 \frac{1}{F_{+}(\xi q)F_{+}(\xi q^{-1})}
 =\sum_{m\geq 0}\chi^{+}_{m}(-\xi)^{m}.
\]
By taking the sum over $\tau=0,1,\ldots,N-1$ 
after the substitution $\xi\mapsto \xi q^{2\tau+1}$, we have
\begin{align}
&\sum_{\tau=0}^{N-1}
 \Big(\prod_{n=1}^{L}\frac{1-t_{p_{n}}^{-N}\xi^{N}}
 {1-t_{p_{n}}^{-1}\xi q^{2\tau}}\Big)
 \frac{1}{F_{+}(\xi q^{2\tau})F_{+}(\xi q^{2\tau+2})}
  \nn\\
&=\sum_{\tau=0}^{N-1}
  \sum_{m\geq 0}\chi^{+}_{m}(-\xi q)^{m}q^{2\tau m}
 =N\sum_{m\geq 0}\chi^{+}_{mN}(-\xi^{N})^{m}
 =N P_{\mathrm{D}}^{+}(\xi^{N}),
  \nn
\end{align}
which proves Proposition~\ref{prop:HW-polynomial}.

We give a remark. One can derive 
Proposition~\ref{prop:HW-polynomial}
from the proof of the spin-1/2 inhomogeneous case 
through the fusion method~\cite{Deguchi_07JPA}. 
However, we have presented the direct 
and straightforward approach here.  


\providecommand{\bysame}{\leavevmode\hbox to3em{\hrulefill}\thinspace}
\providecommand{\MR}{\relax\ifhmode\unskip\space\fi MR }
\providecommand{\MRhref}[2]{%
  \href{http://www.ams.org/mathscinet-getitem?mr=#1}{#2}
}
\providecommand{\href}[2]{#2}

\end{document}